\documentclass[twoside,11pt]{article}

\usepackage[accepted]{melba}
\usepackage{changes}
\usepackage{hyperref}
\usepackage{multicol,lipsum}
\usepackage[shortlabels]{enumitem}
\usepackage{tikz,pgf} 
\usetikzlibrary{arrows,automata} 
\usepackage{verbatim}
\usetikzlibrary{positioning,shapes.geometric}
\usepackage{changes}
\usepackage{amsmath}
\usepackage{algpseudocode}
\usepackage{tabularx}
\usepackage{array}
\usepackage{float}
\usepackage{amsmath}
\usepackage{amssymb}
\usepackage{multirow}
\usepackage{algorithm}
\usepackage{mathtools,xparse}
\DeclarePairedDelimiter{\norm}{\lVert}{\rVert}
\usepackage{amssymb}
\usepackage{gensymb}
\usepackage{array}
\usepackage{tabularx}
\usepackage{enumitem}
\usepackage{lipsum}
\usepackage{color}
\usepackage{graphicx}
\usepackage[caption=false]{subfig}
\usepackage{amsmath}
\usepackage{multirow}
\usepackage[all,poly]{xy}
\usepackage{algpseudocode}
\usepackage{array}
\usepackage[]{siunitx}
\usepackage{textcomp}
\usepackage{nicematrix}
\usepackage[english]{babel}
\usepackage{blindtext}
\usepackage[normalem]{ulem}
\useunder{\uline}{\ul}{}

\usepackage{amsmath,amsfonts}

\def\bfalpha{{\boldsymbol{\alpha}}}

\def\bs0{{\boldsymbol{0}}}

\def\bfx{{\mathbf{x}}}
\def\bfy{{\mathbf{y}}}

\def\bfC{{\mathbf{C}}}
\def\bfD{{\mathbf{D}}}

\def\bfF{{\mathbf{F}}}

\def\bfH{{\mathbf{H}}}

\def\bfL{{\mathbf{L}}}

\def\bfW{{\mathbf{W}}}
\def\bfX{{\mathbf{X}}}
\def\bfY{{\mathbf{Y}}}

\newcounter{algo}
\renewcommand{\thealgo}{\arabic{algo}}

\melbaid{2024:027}  
\doi{https://doi.org/10.59275/j.melba.2024-1656}
\melbaauthors{A. K. Eldaly et al.}  
\volume{2}
\firstpageno{2195}  
\melbayear{2024}  
\datesubmitted{03/2024}  
\datepublished{11/2024}  

\ShortHeadings{Alternative Learning Paradigms for Image Quality Transfer}{A. K. Eldaly et al.}

\title{Alternative Learning Paradigms for Image Quality Transfer}

\author{\firstname Ahmed Karam \surname Eldaly \orcid{0000-0002-2266-6992} \email A.Karam@ucl.ac.uk \\  
	\addr Centre for Medical Image Computing, Department of Computer Science, University College London, UK\\
 Department of Computer Science, University of Exeter, UK
 \AND
	\name Matteo Figini \orcid{0000-0002-8238-2262} \email M.Figini@ucl.ac.uk \\
	\addr Centre for Medical Image Computing, Department of Computer Science, University College London, UK
 	\AND
	\name Daniel C. Alexander \orcid{0000-0003-2439-350X} \email D.Alexander@ucl.ac.uk \\
	\addr Centre for Medical Image Computing, Department of Computer Science, University College London, UK
}

\begin{document}
\maketitle

\begin{abstract}
Image Quality Transfer (IQT) aims to enhance the contrast and resolution of low-quality medical images, e.g. obtained from low-power devices, with rich information learned from higher quality images. In contrast to existing IQT methods in the literature which adopt supervised learning frameworks, in this work, we propose two novel formulations of the IQT problem. The first approach uses an unsupervised learning framework, whereas the second is a combination of both supervised and unsupervised learning. The unsupervised learning approach considers a sparse representation (SRep) and dictionary learning model, which we call IQT-SRep, whereas the combination of supervised and unsupervised learning approach is based on deep dictionary learning (DDL), which we call IQT-DDL. The IQT-SRep approach trains two dictionaries using a sparse representation model using pairs of low- and high-quality volumes. Subsequently, the sparse representation of a low-quality block, in terms of the low-quality dictionary, can be directly used to recover the corresponding high-quality block using the high-quality dictionary. On the other hand, the IQT-DDL approach explicitly learns a high-resolution dictionary to upscale the input volume, while the entire network, including high dictionary generator, is simultaneously optimised to take full advantage of deep learning methods. The two models are evaluated using a low-field magnetic resonance imaging (MRI) application aiming to recover high-quality images akin to those obtained from high-field scanners. Experiments comparing the proposed approaches against state-of-the-art supervised deep learning IQT method (IQT-DL) identify that the two novel formulations of the IQT problem can avoid bias associated with supervised methods when tested using out-of-distribution data that differs from the distribution of the data the model was trained on. This highlights the potential benefit of these novel paradigms for IQT.
\end{abstract}

\begin{keywords}
Image Quality Transfer, Supervised Learning, Unsupervised Learning, Sparse Representation, Dictionary Learning, Deep Dictionary Learning, Deep Learning, Out-of-Distribution, In-distribution
\end{keywords}

\section{Introduction}
Image Quality Transfer (IQT) \cite{alexander2014image, alexander2017image, lin2019deep, tanno2021uncertainty, lin2021generalised, Hongxiang2022, kim20233d} is a machine learning technique that is used to enhance the resolution and contrast of low-quality clinical data using rich information in high-quality images. For example given an image from a standard hospital scanner or rapid acquisition protocol, we might estimate the image we would have got from the same subject using a high-power experimental scanner available only in specialist research centres or a richer acquisition protocols too lengthy to run on every patient. IQT is a vital component of efforts to democratise the capabilities of high power rare experimental systems broadening the accessibility e.g. to lower and middle income countries \cite{anazodo2022framework}. This technique learns mappings from low-quality (e.g. clinical) to high-quality (e.g.experimental) images exploiting the similarity of image structure across subjects, regions, modalities, and scales. The mapping may then operate directly on low-quality images to estimate the corresponding high-quality images. Early work \cite{alexander2017image, blumberg2018deeper, tanno2021uncertainty} focused on diffusion MRI and showed remarkable ability to enhance both contrast and resolution and enabled tractography to recover small pathways impossible to reconstruct at the acquired resolution. Recent work \cite{lin2021generalised} extends the idea to standard structural MRI, particularly targeting application to low-field MRI systems. {IQT technique \cite{alexander2017image} differs from super-resolution in computer vision \cite{lau2023pushing, zhou2020blind, zhou2021image, li2024deep} in several key aspects. In general super-resolution aim to up-sample an image, whereas IQT aims to transfer the quality of information from an image to the other. This means that IQT is not limited to increasing the spatial resolution of images. While super-resolution techniques primarily focus on enhancing the spatial resolution, IQT also aims to improve the image contrast. This dual enhancement is crucial for medical imaging applications where both resolution and contrast are necessary for accurate diagnosis and analysis. Moreover, super-resolution techniques are generally used to upsample images, making them appear sharper and more detailed. In contrast, IQT is specifically designed to transfer the quality from high-quality images to low-quality images. This is particularly beneficial in medical imaging, where high-quality images from advanced scanners are used to enhance the quality of images obtained from lower-power or less advanced scanners. Lastly, IQT differs from modality transfer methods, which maps one modality to another to obtain multi-modality information \cite{iglesias2021joint, iglesias2023synthsr, iglesias2022quantitative}, whereas IQT's primary goal is to enhance the existing image quality, specifically improving resolution and contrast rather than the developing new content. By highlighting these differences, we aim to clearly delineate the unique characteristics and advantages of the IQT task.}

Machine learning models are often trained on a specific data distribution, but may encounter unseen data from different distributions in real-world scenarios. This poses a critical challenge for the security and reliability of machine learning systems, especially in some error-sensitive applications, such as medical diagnosis including the application investigated in this work. One of its powerful capabilities lies in the promising generalisation ability from training data to unseen in-distribution (InD) data. However, the finite training data cannot guarantee the completeness of data distribution, so it is inevitable to encounter out-of-distribution (OOD) data. Machine learning models can be broadly categorised into supervised, unsupervised and self-supervised learning models. In supervised learning, the model is trained by paring inputs with their expected outputs. However, this is far from being practical, since the full data distribution cannot be represented in the training data set. To circumvent this difficulty, unsupervised and self-supervised learning methods can be used.

All IQT models proposed in the literature use supervised learning frameworks to learn a regression between matched patches in low- and high-quality images. In particular deep learning frameworks substantially outperform the original random-forest implementation in terms of global error metrics for enhancement of both diffusion-tensor MRI and low-field structural MRI \cite{alexander2014image, alexander2017image, lin2019deep, tanno2021uncertainty, lin2021generalised, Hongxiang2022}. However, interpretation of images enhanced via such regression models needs caution. First, regression models in general can lead to bias that depends on the training data distribution \cite{obermeyer2019dissecting}. In particular, inputs (here patches) that are rare in the training data are often skewed towards outputs more common in training data; and degenerate regions of the input-space where the mapping is ambiguous are often mapped to a consistent mean giving a false impression of consistent and confident output. Moreover, the performance of deep-learning based methods can degrade even more with OOD data. These effects have been well documented in other image-related regression applications recently, such as parameter mapping \cite{gyori2022training}. So far, they have not been considered in IQT and image enhancement although similar effects are likely to arise.  Additional problems, particularly in deep learning, can arise from over-fitting and under-fitting which can further add to bias in estimates particularly for examples that are over/under-represented in the training data. Moreover, state-of-the-art IQT models, specifically deep neural networks, are generally designed for a static and closed world \cite{krizhevsky2017imagenet, he2015delving}. The models are trained under the assumption that the input distribution at test time will be the same as the training distribution. In real world MRI data, however, deep-learning-based techniques effectiveness diminishes when applied to images that differ significantly from the training data set \cite{gu2019blind}. Although various approaches have been developed to tackle this issue, such as training networks to handle multiple types of degradation \cite{ soh2020meta, xu2020unified, zhang2018learning, zhou2019kernel} and making models less sensitive to degradation through iterative optimisations \cite{shocher2018zero, gu2019blind}, it is also crucial to enhance the robustness of the network structure.

Sparse representation (SRep) using dictionary learning is an unsupervised learning framework that assumes a given signal is sparse in some domain (Wavelets, Fourier, discrete cosine transform, etc.). SRep has proven robust to noise and redundancy in the data, where supervised deep learning algorithms encounter problems \cite{elad2010sparse}. In the IQT context, low and high-quality dictionaries ($\bfD_\ell$, and $\bfD_h$ respectively) can be trained using a sparse representation model using pairs of low- and high-quality volumes. Subsequently, the sparse representation of a low-quality block, in terms of the low-quality dictionary $\bfD_\ell$, can be directly used to recover the corresponding high-quality block using the high-quality dictionary $\bfD_h$. As such, low-quality or high-quality volume patches are represented as a linear combinations of atoms drawn from a dictionary. {SRep has been successfully applied to many other related inverse problems in image processing, such as denoising \cite{li2012group, elad2006image}, restoration \cite{zhang2014group, li2012group}, image quality assessment \cite{liu2017reduced, liu2018blind, liu2024underwater, liu2019unsupervised}, outlier or anomaly detection \cite{eldalyBayesian, eldaly2019patch}, image reconstruction \cite{Eldaly2024Bayesian, Eldaly2024BayesianX}, and super resolution \cite{yang2010image}. In a convex optimisation framework, training and testing samples are forced to follow the observation model of the imaging system on hand. Therefore, any new unseen test samples (either InD or OOD) will follow this model, which can avoid the ``regression to the mean” problems observed with supervised regression models, often observed in OOD data.}

On the other hand, in supervised deep learning, Dong et al. \cite{dong2014learning} replaced the dictionary learning using sparse representation steps described above with a multilayered convolutional neural network to take advantage of the powerful capability of deep learning. As such, the low and high-quality dictionaries are implicitly acquired through network training. Various methods have been proposed to improve the performance of this approach such as in \cite{kim2016deeply, lim2017enhanced, tai2017image, zhang2018image}. However, most of these studies, follow the same formality as in \cite{dong2014learning} from a general perspective, where all the processes in the sparse-coding-based methods are replaced by a multilayered network. Recently, deep dictionary learning \cite{tariyal2016deep} is proposed to take advantage of both transductive and inductive nature of dictionary learning and deep learning, respectively, and is very well suited where there is a scarcity of training data. While dictionary learning focuses on learning ``basis'' and ``features'' by matrix factorisation, deep learning focuses on extracting features via learning ``weights'' or ``filter'' in a greedy layer by layer fashion. Deep dictionary learning has been applied to various problems including recognition \cite{tang2020dictionary, sharma2017deep}, image inpainting \cite{deshpande2020deep}, super resolution \cite{huang2018deep, zhao2017single}, classification \cite{majumdar2017noisy, majumdar2017robust, manjani2017detecting}, and load monitoring \cite{singh2017deep}.

In this work, in contrast to existing IQT models in the literature, we propose two novel IQT algorithms, from which one is an example of unsupervised learning while the other is an example of blended supervised and unsupervised learning. The first approach is based on a sparse representation model and dictionary learning, which we call IQT-SRep. In this approach, low and high-quality dictionaries can be trained using a sparse representation model using pairs of low- and high-quality volumes. Subsequently, the sparse representation of a low-quality block, in terms of the low-quality dictionary, can be directly used to recover the corresponding high-quality block using the high-quality dictionary. The second approach is based on deep dictionary learning which we call IQT-DDL. This approach explicitly learns high-quality dictionary through network training. The main network predicts the high-quality dictionary coefficients, and the weighted sum of the dictionary atoms generates a high-quality output. This approach differs fundamentally from traditional deep-learning methods, which typically employ upsampling layers within the network. The upsampling process in our IQT approach is efficient since pre-generated high-quality dictionary serves as a magnifier during inference. Additionally, the main network no longer needs to retain pixel-level information in the high-quality space, enabling it to focus solely on predicting the dictionary coefficients. The main advantages of these two novel formulations are that they are robust to super resolve heavily OOD test data, and they are well suited where there is a scarcity of training data. We demonstrate the two models using experiments from a low-field MRI application and compare the results with the recently proposed state-of-the-art supervised deep learning approach \cite{Hongxiang2022}. As such, the main contributions of this paper can be summarised as follows.
\begin{enumerate}
    \item We propose two new formulations of the IQT technique, from which one is an unsupervised learning based (IQT-SRep), and one is based on a combination of both supervised and unsupervised learning (IQT-DDL). Both of these formulations have never been previously applied to the IQT problem in literature.
    \item The IQT-SRep approach is based on sparse representation and dictionary learning model and assumes that a given low- or high-quality volume patch can be represented as a linear combination of atoms drawn from a dictionary that is trained using training examples of pairs of low- and high-quality volume patches. This requires training of a pair of coupled dictionaries using a sparse representation model using pairs of low- and high-quality volumes. 
    \item The IQT-DDL approach is based on a combination of supervised and unsupervised learning using deep dictionary learning. This approach assumes that a given low- or high-quality volume patch can be represented as a non-linear combination of atoms drawn from a dictionary that is trained using training examples of pairs of low- and high-quality volume patches. 
    \item We demonstrate the performance of the model using experiments from a low-field MRI application, using both InD and OOD data, and compare with the state-of-the-art supervised deep learning IQT method, for low-field MRI enhancement.
\end{enumerate}

The remaining sections of the paper are organised as follows. Section \ref{sec:ProposedApproach} formulates the problem of IQT using three learning techniques; the formulations that we propose here for IQT-SRep and IQT-DDL are described in detail, and finally, the supervised deep learning approach proposed in \cite{Hongxiang2022} is briefly presented for comparison. Experiments conducted using a low-field MRI application synthesised using data from the human connectome project (HCP) are presented in Section \ref{sec:experiments}. A general discussion is then presented in \ref{sec:Discussion}. Conclusions and future work are finally reported in Section \ref{sec:Conclusion}.

\section{Proposed Approaches}
\label{sec:ProposedApproach}
\subsection{Image quality transfer using sparse representation and dictionary learning (IQT-SRep)}
\subsubsection{Imaging model}
The IQT problem can be mathematically formulated as follows: Given an original vectorised high-quality volume $\bfX \in \mathbb{R}^{M}$, its corresponding low-quality version is denoted as $\bfY \in \mathbb{R}^{P}$, where the relation between the two volumes can be modeled as
\begin{equation}
    \begin{aligned}
    \bfY = \bfL\bfH\bfX + \bfW,
    \label{eq:Model0}
    \end{aligned}
\end{equation}
where $\bfH$ is the matrix representing a linear blurring operator, $\bfL$ is the downsampling operator, and $\bfW$ stands for additive noise, modelling observation noise and model mismatch and is assumed to be a white Gaussian noise sequence. This equation states that $\bfY$ is a blurred and down-sampled version of the original high-quality volume $\bfX$.

In IQT, the goal is to recover a high-quality volume $\hat{\bfX}$ given its blurred and down-sampled version $\bfY$, such that $\hat{\bfX} \approx \bfX$. The problem of estimating $\bfX$ from $\bfY$ in Eq. \eqref{eq:Model0} is an ill-posed linear inverse problem (LIP), i.e., the matrix $\bfL\bfH$ is singular and/or very ill-conditioned, since for a given low-quality input, infinitely many high-quality volumes satisfy the above equation. Consequently, this problem requires additional regularisation (or prior information from Bayesian perspective) in order to reduce uncertainties and improve estimation performance.

\begin{figure}
	\centering
		\includegraphics[width=0.85\textwidth]{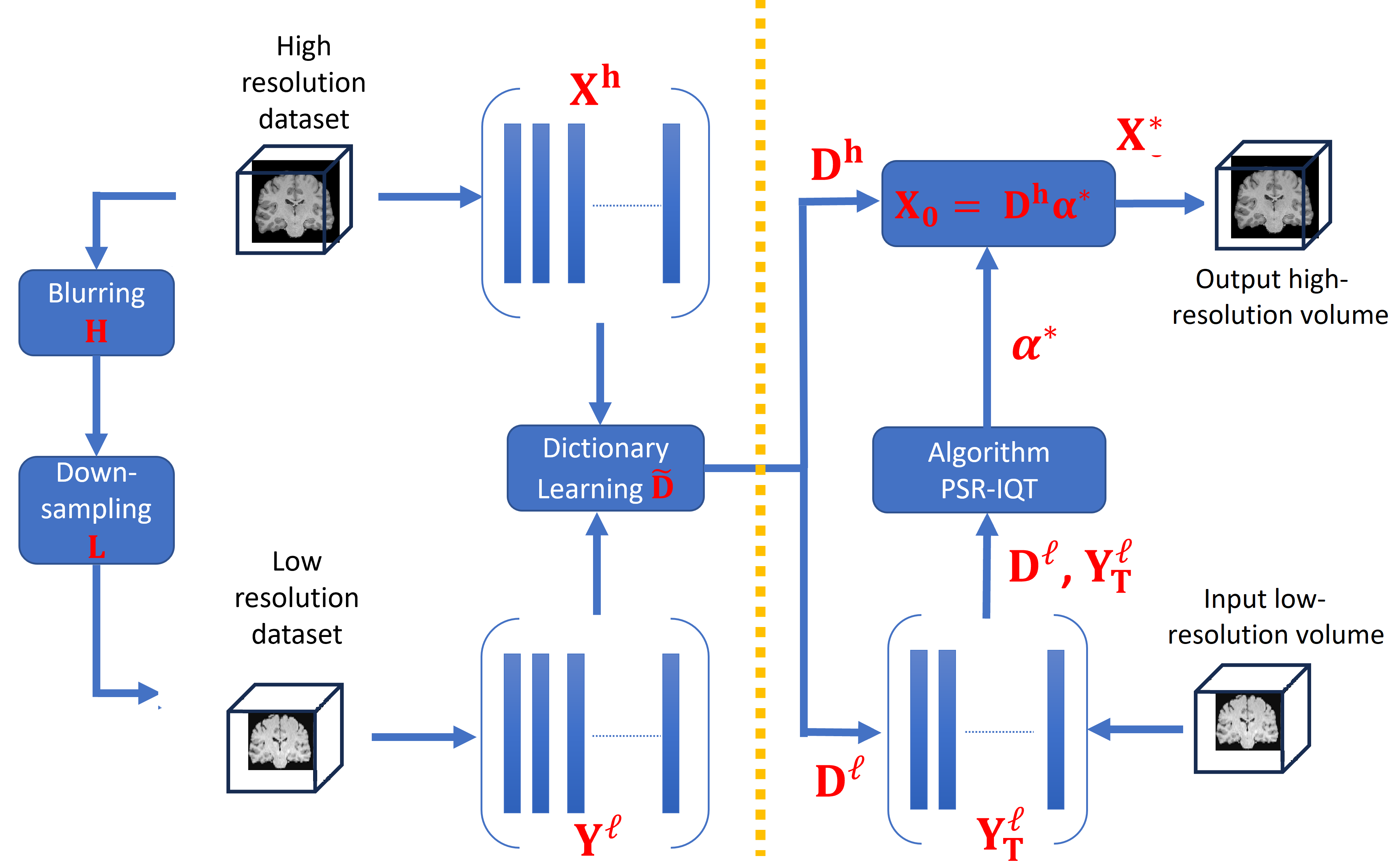}
	\caption{{A schematic diagram of the proposed IQT approach using sparse representation and dictionary learning (IQT-SRep), where $\bfD_h$: High-quality dictionary, $\bfD_\ell$: Low-quality dictionary, $\bfY$: Low-quality input volume, $\bfX_0$: Initial high-quality volume, $\lambda, \beta$ Regularisation parameters, $\sqrt[3]{m}$: Patch size, $p$: Number of pixel overlap, $s$: Scale, $\bfy$: A patch from the low-quality image $\bfY$, $\mu$: Mean intensity of the patch \(\bfy\), $\bfalpha$: Sparse representation coefficients, $\bfalpha^*$: Optimised sparse representation coefficients, $\bfF$: Transformation matrix, $\bfx$: High-quality patch, and $\bfX^*$: High-quality volume. }}
	\label{fig:algo}
\end{figure}

Figure \ref{fig:algo} shows a schematic diagram to the IQT problem using a sparse representation model and dictionary learning. The proposed model consists of two separate stages. First, the coupled low-quality and high-quality dictionaries, $\bfD_\ell$ and $\bfD_h$ respectively, are constructed from training data set. Then, a reconstruction algorithm is applied to upscale a test low-quality volume to recover its high-quality version. This algorithm considers the patch-based sparse prior model to recover an estimate to the high-quality volume in a patch-by-patch basis. The following sections provide more details about the two stages mentioned above.

\subsubsection{Joint dictionary construction}
\label{subsec:DictionaryLearning}
Constructing the high-quality and low-quality dictionaries requires a set of matched high- and low-quality volume patches. The training set is composed by a set of high-quality and the corresponding low-quality volumes. As proposed by \cite{zeyde2010single}, the high-quality volumes are processed to obtain only the high-frequency information, whereas the intensity maps are used for the low-quality volumes. Each of the high- and low-quality volumes are then split into a set of 3D patches which are vectorised and training pairs are generated. Patches containing $>80\%$ background voxels are excluded from the patch library. The coupled-dictionary training algorithm proposed by \cite{zeyde2010single} is then used in order to obtain the low- and high-quality dictionaries $\bfD_\ell$ and $\bfD_h$ respectively. For this local model, the two dictionaries $\bfD_h$ and $\bfD_\ell$ are trained such that they share the same sparse representations for each high- and low-quality volume patch pair. Finally, the dimensionality of ${\bfD_\ell}$ may be reduced to speed up the subsequent computations, given the intrinsic redundancy of the multi-scale edge analysis. For doing so, a Principal Component Analysis (PCA) is applied to this matrix, searching for a set of projection coefficients that represents at least $90\%$ of the original variance. All patches are collected together to form the reduced low-quality dictionary $\bfD_\ell$, whereby the number of atoms in the dictionary has not changed.

\subsubsection{Patch-based sparsity prior model}
The low-quality volume $\bfY$ can be split into a set of overlapping 3D patches $\bfy$, each of size $\sqrt[3]{m}\times \sqrt[3]{m}\times \sqrt[3]{m}$. With the sparse generative model, each patch $\bfy$ can be represented by a linear combination of a few atoms drawn from a dictionary $\bfD_\ell$, which characterises the low-quality patches. This can be written as
\begin{equation}
    \begin{aligned}
    \bfy = \bfD_\ell\bfalpha_\ell,
    \label{eq:sparsity0}
    \end{aligned}
\end{equation}
where \text{$\bfalpha \in \mathbb{R}^K$} is a sparse vector and $\norm{\bfalpha}_0 \ll K$. The corresponding high-quality patch $\bfx$, with size $\sqrt[3]{p}\times \sqrt[3]{p}\times \sqrt[3]{p}$, can be computed by again applying the following sparse generative model
\begin{equation}
    \begin{aligned}
    \bfx = \bfD_h\bfalpha_h.
    \label{eq:sparsity01}
    \end{aligned}
\end{equation}

From Eq. \eqref{eq:sparsity0} and \eqref{eq:sparsity01}, it can be assumed that the sparse representation of a low-quality patch in terms of $\bfD_\ell$ can be directly used to recover the corresponding high-quality patch from $\bfD_h$, namely, that $\bfalpha_\ell = \bfalpha_h$. Therefore, the reconstructed high-quality image $\hat{\bfX}$ can be built by applying the sparse representation to each patch $\bfy$ in $\bfY$ and then using the estimated $\bfalpha$ with $\bfD_h$ to obtain each $\bfx$, which together form the image $\hat{\bfX}$. 

\subsubsection{Local reconstruction by sparsity}
The aim is to estimate a high-quality version $\tilde{\bfX}$ from a given low-quality volume $\bfY$. Given a test low-quality volume, for each input low-quality patch $\bfy$, we find a sparse representation with respect to $\bfD_\ell$. The corresponding high-quality patch bases $\bfD_h$ will be combined according to these coefficients to generate the output high-quality patch $\bfx$. The problem of finding the sparsest representation of $\bfy$ can be formulated as
\begin{equation}
\begin{aligned}
\underset{\bfalpha}{\text{minimise}} & \hspace{0.3cm} \frac{1}{2}\norm{\bfF\bfD_\ell\bfalpha - \bfF\bfy}_2^2 + \lambda\norm{\bfalpha}_1,
\label{eq:sparsity2}
\end{aligned}
\end{equation}
where $\lambda$ balances sparsity of the solution and fidelity of the approximation to $\bfy$, and $\bfF$ is a linear feature extraction operator as in \cite{zeyde2010single}. Given the optimal solution $\bfalpha^*$ of Eq.\eqref{eq:sparsity2}, the high-quality patch $\bfx$ can be reconstructed as $\bfx = \bfD_h\bfalpha^*$. This optimisation problem can be solved using the Basis Pursuit algorithm \cite{chen1994basis}.

{The complete IQT process is summarised in Algorithm (\ref{algo:PSRIQT}). In this algorithm, the input low-quality volume $\bfY$ is up-sampled using bicubic interpolation to provide a preliminary high-resolution volume. For each cubic patch of size $\sqrt[3]{m} \times \sqrt[3]{m} \times \sqrt[3]{m}$ from the up-sampled volume, starting from the top left corner with an overlap $p$, the mean intensity $\mu$ of the patch is computed to ensure that the dictionary represents image textures rather than absolute intensities. The sparse representation $\bfalpha^*$ of the patch is then obtained by solving an optimisation problem that minimises the difference between the transformed low-quality patch and its sparse representation in the low-quality dictionary $\bfD_\ell$, subject to a sparsity constraint controlled by the regularisation parameter $\lambda$. Using the high-quality dictionary $\bfD_h$ and the sparse coefficients $\bfalpha^*$, a high-quality patch $\bfx$ is generated. This high-quality patch, with the mean intensity restored, is placed in the corresponding location in the initial high-quality volume $\bfX_0$. After processing all patches, the final high-quality volume $\bfX$ is obtained.}

\begin{algorithm}
\caption{IQT using patch-based sparse representation and dictionary learning (IQT-SRep)}
\label{algo:PSRIQT}
\begin{algorithmic}[1]
\item \textbf{Input}: $\bfD_h, \bfD_\ell$ and $\bfY$
\item \textbf{Initialise} $\bfX_0$, \textbf{Choose} Regularisation parameters $\lambda, \beta,$ Patch-size $\sqrt[3]{m}$, pixel-overlap $p$ and scale $s$
\item Up sample the input low-quality volume using bicubic interpolation.
\item \textbf{For} each $\sqrt[3]{m}\times \sqrt[3]{m}\times \sqrt[3]{m}$ patch $\bfy$ from an image $\bfY$, from top left corner of the volume, with an overlap $p$
\begin{itemize}
    \item {Compute:} mean intensity $\mu$ of the patch $\bfy$
    \item {Solve:} $\bfalpha^* = \underset{\bfalpha}{\text{minimise}} \hspace{0.3cm} \frac{1}{2}\norm{\bfF\bfD_\ell\bfalpha - \bfF\bfy}_2^2 + \lambda\norm{\bfalpha}_1$
    \item Generate the high-quality patch $\bfx = \bfD_h\bfalpha^*$
    \item Place the high-quality patch $\bfx + \mu$ in the high-quality volume $\bfX_0$
\end{itemize}
\item \textbf{End}
\item \textbf{Output} High-quality volume $\bfX = \bfX_0$
\end{algorithmic}
\end{algorithm}

\subsection{Image quality transfer using deep dictionary learning (IQT-DDL)}
\label{subsec:DeepDictionaryProposedApproach}
\begin{figure}
    \centering
    \includegraphics[width=0.99\textwidth]{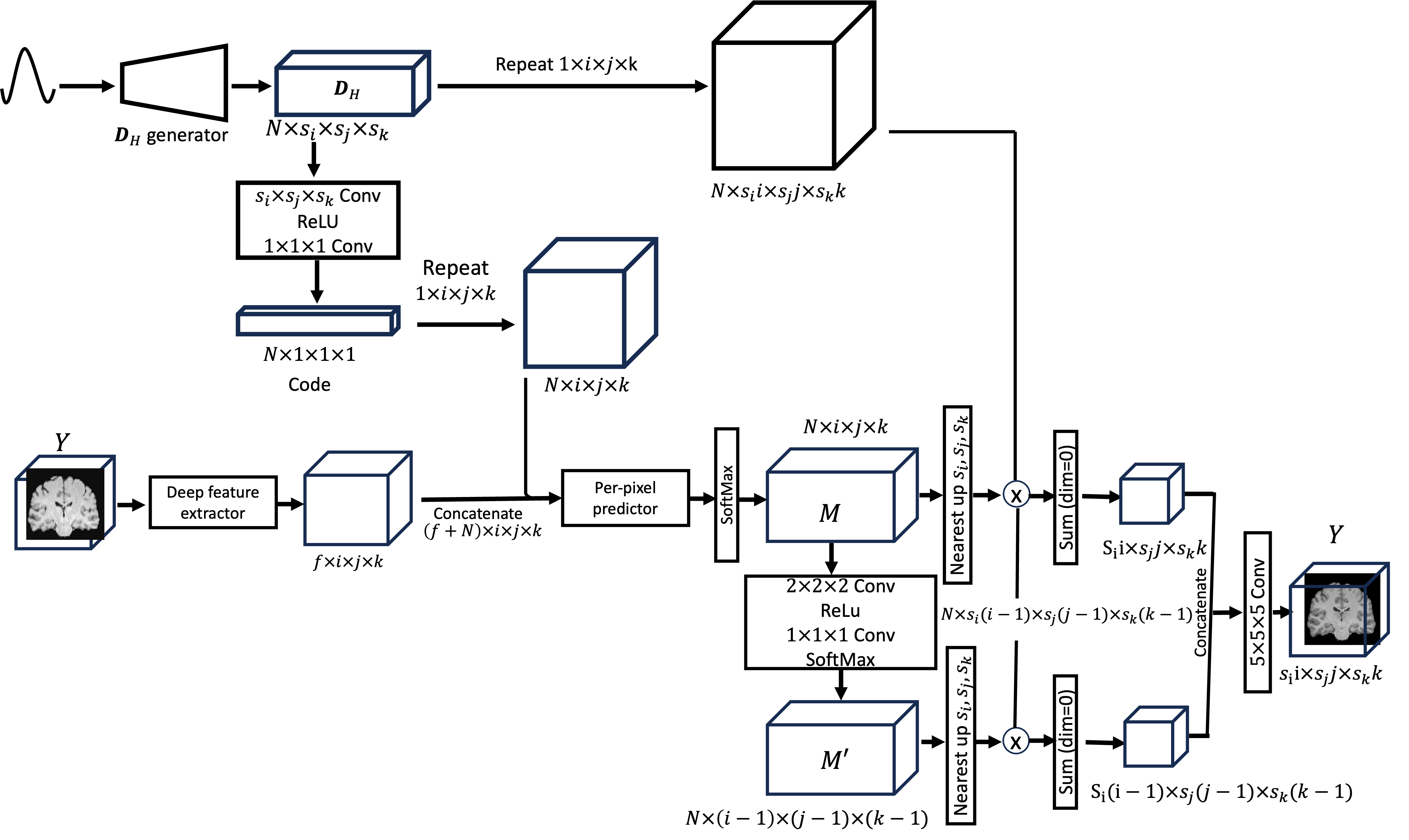}
    \caption{A schematic diagram of IQT using deep dictionary learning. Random noise generates the high-resolution dictionary $\bfD_H$. Then a per-pixel predictor takes as input a concatenation of an encoded code of $\bfD_H$ and an extracted feature. A final image  based on $\bfD_H$ is then constructed using predictor output.}
    \label{fig:model}
\end{figure}

The IQT using a deep dictionary learning model is composed of three main steps: constructing the high-quality dictionary $\bfD_H$, per-pixel prediction, and finally image reconstruction from patches. The high-quality dictionary $\bfD_H$ is generated from random noise input. The per-pixel predictor then estimates the coefficients of $\bfD_H$ for each pixel from a low-quality input. In the reconstruction phase, the high-quality image can be computed using the weighted sum of the elements (or atoms) of $\bfD_H$. In this work, we use L1 loss function to optimise the network $L = \frac{1}{M}\sum_{m = 1}^{M}\norm{I_m^{gt} - \Theta(I_m^{lq})}_1,$ where $I_m^{lq}$ and $I_m^{gt}$ are low- and high-quality patches respectively, $M$ is the number of training pairs, and $\Theta(\cdot)$ represents a function of the IQT-DDL network. Figure \ref{fig:model} provides a schematic diagram of the proposed method. The following sections provide more details about each step.

\subsubsection{Construction of the high-quality dictionary $\bfD_H$} 
The high-quality dictionary $\bfD_H^{(N\times s_i\times s_j\times s_k)}$ is constructed from random noise using a standard Gaussian distribution, where $s_i, s_j$ and $s_k$ are up-scaling factors in $i$, $j$ and $k$ directions, and $N$ is the number of dictionary atoms. The high-quality dictionary $\bfD_H$ is then encoded by $s_i\times s_j\times s_k$ convolution with groups $N$, followed by ReLU \cite{nair2010rectified} and $1\times 1\times 1$ convolution. Each $N$ element of the resultant code $\bfC_H^{N\times 1\times 1\times 1}$ represents each $s_i\times s_j\times s_k$ atom as a scalar value. Note that low-quality dictionaries can be naturally replaced by convolutional operations, and therefore only $\bfD_H$ is constructed. The $\bfD_H$ generator has a tree-like structure, where the nodes consist of two $1 \times 1\times 1$ convolutional layers with ReLU activation. The final layer has a Tanh activation followed by a pixel shuffling layer. To produce $N$ atoms, depth $d$ of the generator is determined as $d = \log_2(N)$. 

\subsubsection{Per-pixel Prediction}
We use the UNet++ \cite{zhou2018unet++} as a deep feature extractor in Fig. \ref{fig:model}, with depth of three, and a long skip connection is added. For an input image $I \in \mathbb{R}^{i\times j\times k}$, the deep feature extractor generates a tensor of size $f \times i \times j\times k$. The per-pixel predictor then takes as input a concatenation of the extracted feature and the expanded code of $\bfD_H$, such that $C_H^{N\times i\times j\times k} = R_{1\times i\times j\times k}(C_H^{N\times 1\times 1\times 1})$, where $R_{a \times b \times c(\cdot)}$ denotes the $a \times b \times c$ repeat operations. The per-pixel predictor is composed of ten bottleneck residual blocks followed by a softmax function that computes the $N$ coefficients of $\bfD_H$ for each input pixel. Both the deep feature extractor and per-pixel predictor contain batch normalisation layers \cite{loffe2014accelerating} before the ReLU activation. The resultant prediction map $M^{N\times i\times j\times k}$ is further convolved with a $2 \times 2\times 2$ convolution layer to produce a complementary prediction map $M'^{N \times (i-1)\times (j-1)\times (k-1)}$, that compensates the patch boundaries when reconstructing the final output. The detail of the compensation mechanism is described in the next subsection.

\subsubsection{Reconstruction}
The prediction map $M^{N\times i\times j\times k}$ is upscaled to $N \times s_ii\times s_jj \times s_kk$ by nearest-neighbor interpolation, and the element-wise multiplication of that upscaled prediction map $U_{s_is_js_k}(M^{N\times i\times j\times k})$ with the expanded dictionary $R_{1\times i\times j\times k}$ $(D_H^{N\times s_i\times s_j\times s_k})$ produces $N \times s_ii\times s_jj\times s_kk$ tensor $T$ consists of weighted atoms. The $U_{s_i,s_j,s_k}(\cdot)$ denotes $s_i\times s_j\times s_k$ nearest neighbor upsampling. Finally, the tensor $T$ is summed over the first dimension, producing the output $x$ as

\begin{equation}
    \begin{aligned}
    x^{1\times s_ii\times s_jj\times s_kk} = \sum_{n=0}^{N-1}T^{N\times s_ii\times s_jj\times s_kk}[n, :, :, :],
    \end{aligned}
\end{equation}

\begin{equation}
    \begin{aligned}
    T^{N\times s_ii\times s_jj\times s_kk} = U_{s_i,s_j,s_k}(M^{N\times i\times j\times k})\otimes R_{1\times i\times j\times k}(D_H^{N\times s_i\times s_j\times s_k}).
    \end{aligned}
\end{equation}

The same sequence of operations is applied to the complementary prediction map to obtain the output $x'$. The final high-field prediction is obtained by centering $x$ and $x'$ on top of each other and concatenating the overlapping parts of the centered $x$ and $x'$, and applying a $5 \times 5\times 5$ convolution. For non-overlapping parts, $x$ is simply used as the final output.

\subsection{Image quality transfer using deep learning (IQT-DL)}
\label{subsec:DLTraining}
A supervised learning IQT algorithm which was implemented using a deep learning framework (IQT-DL) is recently proposed \cite{Hongxiang2022, lin2019deep}. This approach was used for IQT application in low-field MRI and showed superior performance compared to existing methods. The model is based on an anistropic U-Net trained on matched pairs of image patches from real high-field and synthetic low-field volumes generated by a stochastic decimation model which is presented in the Experiments section. This model considered the anisotropic U-Net architecture, which is an adaptation of the U-Net architecture to map input and output patches that differ in voxel dimension by the downsampling factor, $s$, in the slice direction. The main additions to the classic U-Net architecture are a bottleneck block, connecting corresponding levels of the contracting and expanding paths, and a residual core used to include more convolutional layers on each level. All convolution layers are activated by Rectified Linear Unit (ReLU) with Batch Normalisation (BN). The average voxel-wise mean square error over all patch pairs was used as a loss function. {For more details and a block diagram of their proposed approach, see \cite{Hongxiang2022}.}

\section{Experiments}
\label{sec:experiments}
\begin{figure}[ht]
\centering
\includegraphics[width=0.75\textwidth]{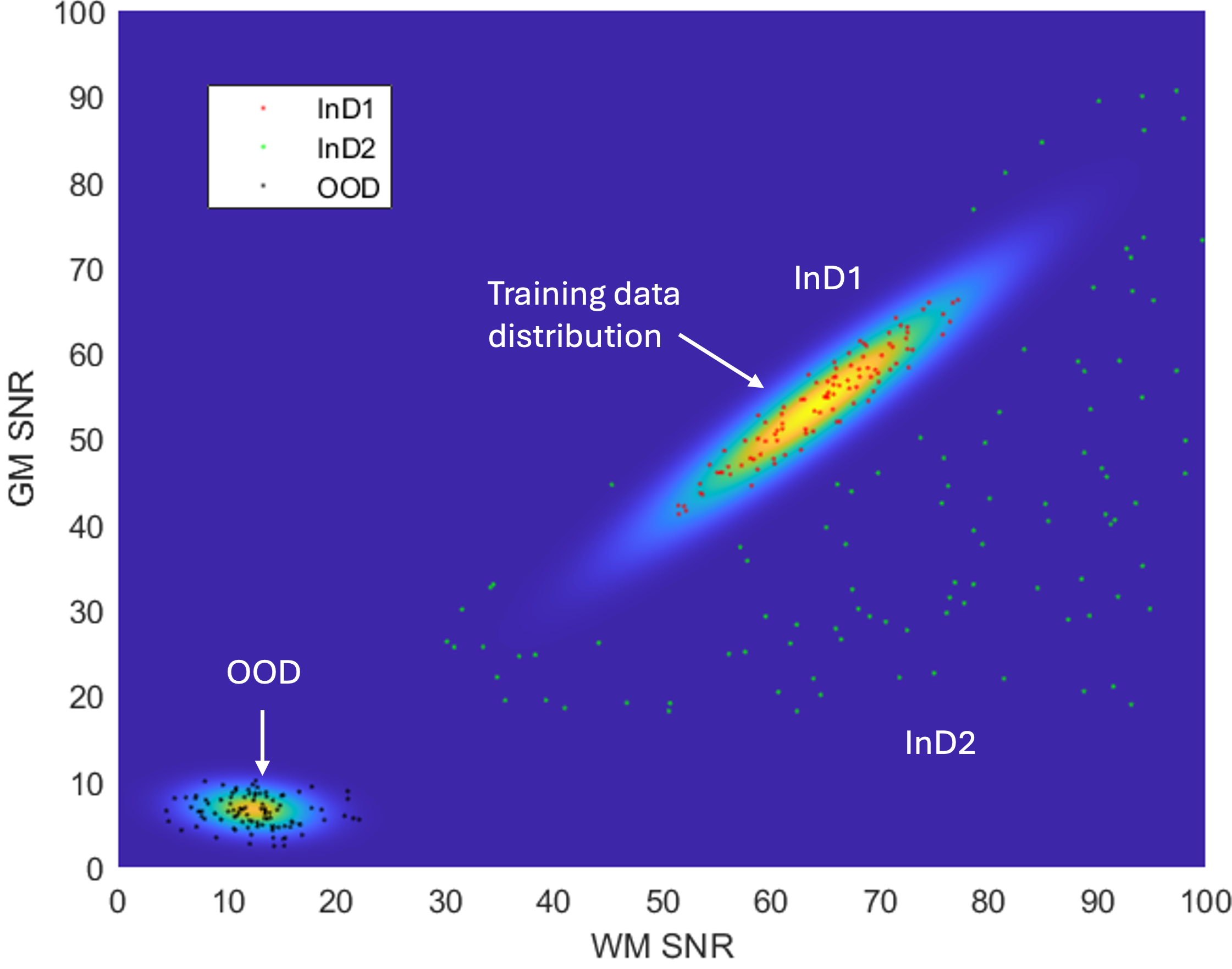}
\caption{A schematic diagram of training and testing datasets. The two test in distribution datasets (InD1 and InD2) are synthesised using parameters sampled from the same 2D Gaussian distribution used for the training set. In particular, InD1 is synthesised with parameters using a Mahalanobis distance $< 1$, and InD2 with Mahalanobis distance $> 3$, with the constraint of having the SNR higher in WM than in GM, to keep the tissue contrast compatible with T1w. The out-of-distribution (OOD) data set is simulated using parameters sampled from a distribution estimated from ultra-low field T1w images.}
\label{fig:InDOOD}
\end{figure}

\vspace{-0.2cm}
The performance of the proposed IQT-SRep and IQT-DDL approaches is demonstrated using a low-field MRI application, using both in-distribution (InD) and out-of distribution (OOD) datasets. The aim is to recover contrast enhanced and super-resolved images akin to those obtained using high field MRI scanners, standard in higher income countries, from low-field MR images form scanners still widely used in low-and-middle class income countries (LMICs). The proposed approaches are compared against the state-of-the-art supervised deep learning framework (IQT-DL) \cite{Hongxiang2022, lin2019deep}, described in the previous section, to reveal both advantages and disadvantages of each of them. {The main data set for training and testing is derived from the T1-weighted MRI images provided by the Human Connectome Projects (HCP), acquired on a 3 Tesla Siemens Connectome scanner \cite{Sotiropoulos2013}, with a $0.7$-mm isotropic voxel. The repetition time (TR), echo time (TE), and inversion time (TI) for T1w are set to $2400, 2.14, \text{and }1000$ ms, respectively. We have chosen 65 subjects, from which 60 were used for training and 5 for testing. {The training and testing datasets are synthesised using a stochastic low-field simulator described in \cite{Hongxiang2022}, the inputs of which are the signal-to-noise ratio (SNR) in gray matter (GM) and white matter (WM). The training data set is built using, for each synthetic volume, a randomly sampled SNR pair from the bivariate Gaussian distribution estimated from a real low-field MRI data set acquired in Nigeria \cite{Hongxiang2022}. Three Low-field test datasets, five volumes each, are synthesised. Two test datasets are synthesised using parameters sampled from the same 2D Gaussian distribution used for the training set, and are called in-distribution data (InD1 and InD2). In particular, InD1 is synthesised with parameters using a Mahalanobis distance $< 1$, and InD2 with Mahalanobis distance $> 3$, with the constraint of having the SNR higher in WM than in GM, to keep the tissue contrast compatible with T1w. The simulation parameters of third data set are sampled from a distribution estimated from ultra-low field T1w images, and is called out-of-distribution (OOD) data set. Figure \ref{fig:InDOOD} shows a schematic diagram of both training and testing data structure, with the stochastic low-field image simulator for training and testing samples described below.}}
\subsection{Model training for IQT-SRep, IQT-DDL, and IQT-DL}
Once the training set of matched low-and high-field pairs is composed as explained above, paired patches are obtained by cropping corresponding high-quality and synthetic low-quality volumes into patches at regularly spaced locations. Patches containing $>80\%$ background voxels are excluded from the patch library. Training details of the IQT-SRep, IQT-DDL and IQT-DL models are presented below.
\subsubsection{IQT-SRep}
The number of atoms and patch-sizes in dictionaries $\bfD_\ell$ and $\bfD_h$ has impact on two important aspects of the proposed IQT-SRep model; that are the reconstruction accuracy and reconstruction time. Larger dictionaries include more image patterns, and therefore more accurate super-resolved volumes. However, the drawbacks are the computational complexity of solving the optimisation problem and the longer time required for patch extraction. Following this, from an initial set of $100,000$ 3D-vectorised patches, we learned compact dictionaries of different atom numbers, including 150, 256, 512, 1024 and patch-sizes of $3\times 3\times 3$, $5\times 5\times 5$ and $7\times 7\times 7$. We first present those of $1024$ atoms using $7\times 7\times 7$ patch-size, which provide best construction quality, and the effect of different atom number is presented afterwards.
\subsubsection{IQT-DDL}
In this work, we adopt a model using different atoms numbers of 64 and 128 atoms. The number of filters of the models is adjusted according to the number of atoms. The scaling factors $s_i, s_j, \text{ and } s_k$ are set to $s_i = 1, s_j = 1, \text{ and } s_k = 4$. The network is trained using low-quality patch size of $32\times 32\times (32/s_k)$ with a mini-batch size of 32. Random flipping and rotation augmentation is applied to each training sample. An Adam optimiser \cite{kingma2014adam} with $\beta_1 = 0.9, \beta_2 = 0.999,$ and $\epsilon = 10^{-8}$ is used. The learning rate of the network except for the $\bfD_H$ generator is initialised as $2e^{-4}$ and halved at $[200k, 300k, 350k, 375k]$. The total training iterations is $400k$. The learning rate of the $\bfD_H$ generator is initialised as $5e^{-3}$ and halved at $[50k, 100k, 200k, 300k, 350k]$. In addition, to stabilise training of the $\bfD_H$ generator, we randomly shuffle the order of output atoms for the first $1k$ iterations. The results of the 128 atoms dictionary are first presented, followed by a comparison with those of 64 atoms dictionary. 
\subsubsection{IQT-DL}
As in \cite{Hongxiang2022}, we use a default patch size of $32\times 32\times (32/s_k)$ and $32\times 32\times 32$, respectively for low-field and high-field volumes, and a step size of $8$, $16$, and $16/s_k$ along $x$-, $y$-, and $z$-directions, which provide best construction quality. Training model is then constructed using the training procedure explained in Section \ref{subsec:DLTraining}.
\subsection{Testing}
Each test volume is split into overlapping patches of size similar to that used for training in each model. The trained IQT-SRep, IQT-DDL and IQT-DL models described above are then applied to each of these patches to estimated the high-field volumes. The magnification factor $s_k$ for all models is set to $4$. For the IQT-SRep model, in all experiments, the $\lambda$ parameter is set to $0.01$. Slight variation of this parameter does not change the results significantly.
\subsection{Evaluation}
The quantitative measure used to assess the quality of the IQT algorithms presented in the previous section are the normalised root mean squared error (NRMSE), defined as
\begin{equation}
    \begin{aligned}
        \text{NRMSE} = \frac{\sqrt{\frac{\sum_{n=1}^{N} (\bfx_n - \hat{\bfx_n})^2}{N}}}{\text{Max}(\bfx)},
    \end{aligned}
\end{equation}
where  $\bfx$ is the ground truth high-quality image, $\hat{\bfx}$ is the corresponding estimate from the low-field counterpart, and $\text{Max}(\bfx)$ is the maximum intensity of the ground truth high-field image $\bfx$, {and structural similarity index measure (SSIM) which can be computed as in \cite{wang2004image}.}
\subsection{Results}
We utilise the proposed unsupervised learning IQT-SRep, the supervised deep learning IQT-DL and the blended learning IQT-DDL approaches to super resolve the testing datasets InD1, InD2 and OOD described above. Below, we show the quantitative and the qualitative performance, as well as the effect of changing different crucial parameters such as atom number in IQT-SRep and IQT-DDL approaches. 
\subsubsection{Quantitative results}
Table \ref{tab:NRMSE} provides NRMSE and SSIM results of InD1, InD2 and OOD using the three methods IQT-SRep, IQT-DDL and IQT-DL. We can observe that the supervised deep learning approach IQT-DL provides better results (lowest NRMSE and highest SSIM) using the in-distribution datasets (InD1 and InD2), compared to the unsupervised learning IQT-SRep algorithm, revealing that supervised learning is more robust for super-resolving images that follow the same distribution of the training data set compared to unsupervised learning. However, when testing using out-of distribution data that is different from the distribution of the training samples, the unsupervised learning approach IQT-SRep provides lower NRMSE and higher SSIM compared to the supervised deep learning model IQT-DL. This highlights the importance of unsupervised learning models since the full data distribution cannot be represented in the training data set. On the other hand, we can observe that the supervised deep learning model IQT-DL performs better (lower NRMSE and higher SSIM) than the blended supervised and unsupervised learning IQT-DDL approach using InD1, whereas the IQT-DDL provides better results using both InD2 and OOD datasets. This reveals the robustness of the blended learning IQT-DDL approach in super-resolving datasets differ from that the model was trained on, in addition to data that slightly deviates from InD1 but still part of the training samples.

\subsubsection{Qualitative results}
\begin{figure}
	\centering
		\includegraphics[width=0.83\textwidth]{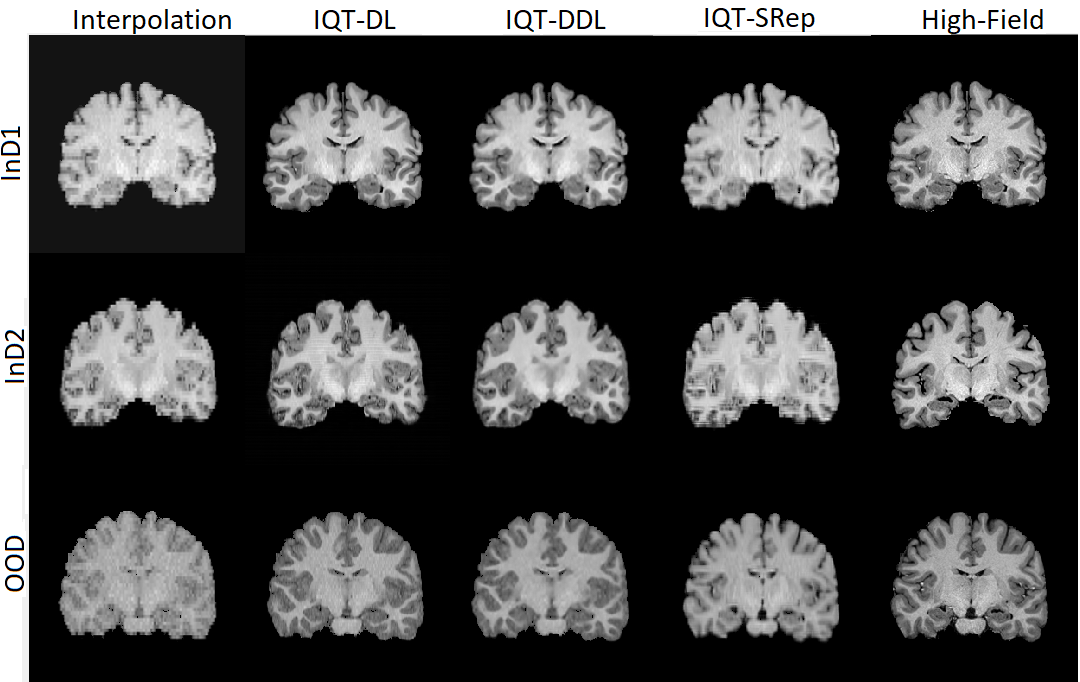}
	\caption{Results using the HCP data set on coronal direction of the three different data distributions InD1, InD2 and OOD (rows) using IQT-SRep, IQT-DDL and IQT-DL. First column shows interpolated low-field image, second to forth columns show image estimate using IQT-DL, IQT-DDL and IQT-SRep, respectively, and fifth column shows original high-field image.}
	\label{fig:Estimates}
\end{figure}

\begin{figure}
	\centering
		\includegraphics[width=0.85\textwidth]{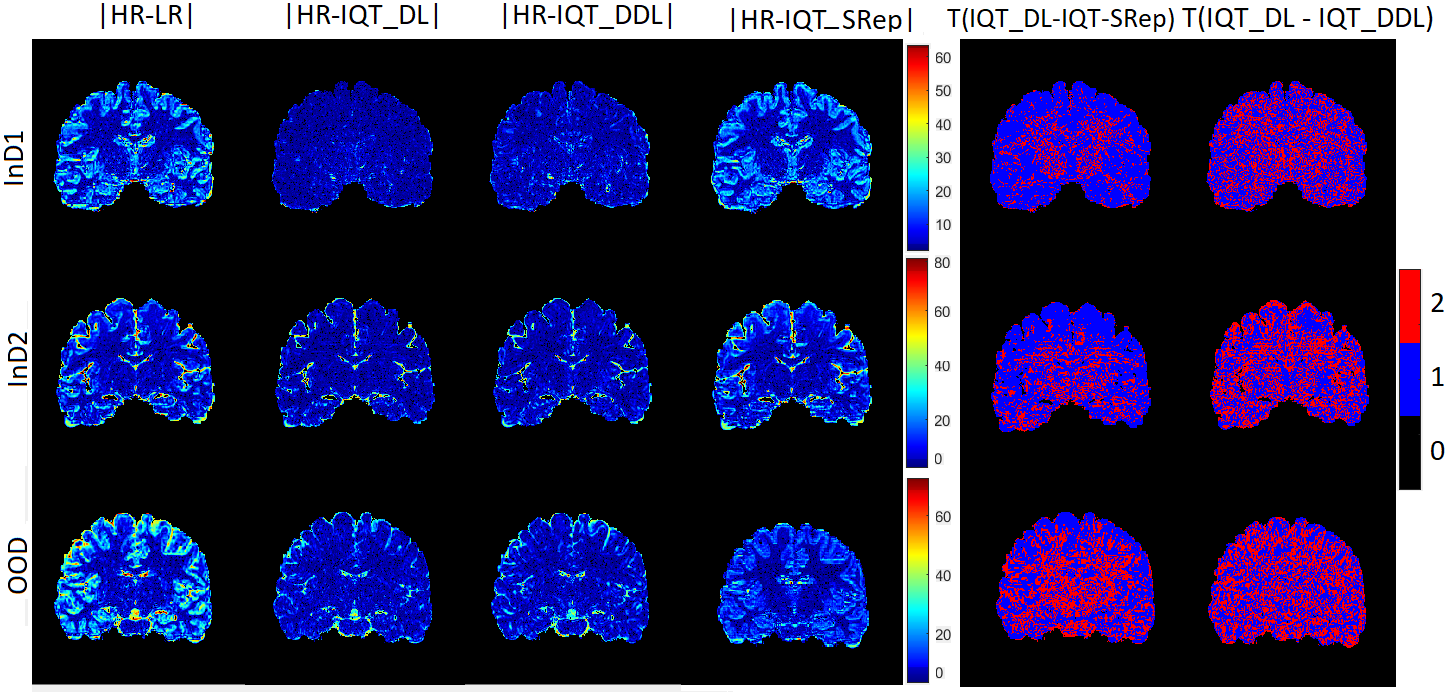}
	\caption{Absolute errors for results in Figure \ref{fig:Estimates}, between gold-standard high-field image (Column 5 of Figure 4), and Column 1: corresponding low-quality image, Column 2: IQT-DL, Column 3: IQT-DDL, and Column 4: IQT-SRep. Columns 5 and 6 show binary maps of regions (in red label) where the IQT-SRep and IQT-DDL, respectively provide closer estimates to the gold-standard high-field images compared to IQT-DL.}
	\label{fig:Errors}
\end{figure}

Figure \ref{fig:Estimates} shows examples of coronal T1 weighted images from the HCP data set, corresponding to synthesised low-field images using InD1, InD2 and OOD, and results of IQT-SRep, IQT-DDL and IQT-DL. Figure \ref{fig:Errors} shows corresponding absolute error maps between high-quality ground truth images and corresponding low-quality images, and results of IQT-SRep, IQT-DDL and IQT-DL. Moreover, the binary maps of regions (in red label) where the IQT-SRep and IQT-DDL models provide closer estimates to ground truth high-quality images compared to IQT-DL are also presented. The qualitative results in general follow the same behaviour of the quantitative results described earlier: although IQT-DL provides better visual results of brain structure compared to IQT-SRep using InD1, and InD2, the IQT-SRep model shows better visual results using OOD data compared to IQT-DL. This is clearer in the absolute error maps in Fig. \ref{fig:Errors}, between high-quality ground truth images and results of both IQT-DL and IQT-SRep, and in the binary maps where there are more image regions where IQT-SRep and IQT-DDL performs better than IQT-DL. This implies that the IQT-SRep approach is more robust for image enhancement using out-of-distribution data, which are created using a different distribution to that of the training samples mimicking real-world examples. Moreover, the IQT-SRep approach provides smoother outputs compared to that of the IQT-DL approach where artifacts arising from patch construction are very obvious. On the other hand, the blended learning IQT-DDL approach provides better visual results than the supervised deep learning approach IQT-DL using InD2 and OOD datasets. This is also clear in the absolute error maps and in the binary maps where there are more image regions where IQT-DDL performs better than IQT-DL in Fig. \ref{fig:Errors}. {On the other hand, while in this work we process data volumes by splitting them into overlapping patches, the proposed approaches ensure that information from the borders of each patch is preserved and integrated into the subsequent patches, thereby there is no information loss and the continuity of image features across the entire volume is maintained. Moreover, we synthesise the low-quality volumes from the high-quality ones, which ensures that there are no pixel alignment problems, as both low-quality and high-quality volume pairs are inherently aligned during the synthesis.}

To summarise, the blended learning IQT-DDL approach provides best visual results compared to the supervised deep learning IQT-DL and the unsupervised leaning IQT-SRep approaches using both InD2 and OOD datasets, whereas the unsupervised learning approach IQT-SRep provides better visual results than the supervised deep learning IQT-DL approach using OOD which is generated using a different distribution to that of the training samples. There are widespread regions where the errors are lower for IQT-SRep and IQT-DDL compared to IQT-DL highlighting bias in the regression model estimates, which both IQT-SRep and IQT-DDL can avoid. 
\begin{table}[]
\centering
\caption{{Normalised root mean squared error (NRMSE), and structural similarity index measure (SSIM) using in-distribution data (InD1), and (InD2), and out-of-distribution data (OOD). Best results are highlighted in bold font, and second best are underlined.}}
\label{tab:NRMSE}
\begin{tabular}{c|cc|cc|cc|cc|}
\cline{2-9}
 & \multicolumn{2}{c|}{\textbf{Interpolation}} & \multicolumn{2}{c|}{\textbf{IQT-SRep}} & \multicolumn{2}{c|}{\textbf{IQT-DDL}} & \multicolumn{2}{c|}{\textbf{IQT-DL}} \\ \cline{2-9} 
 & \multicolumn{1}{c|}{NRMSE} & SSIM & \multicolumn{1}{c|}{NRMSE} & SSIM & \multicolumn{1}{c|}{NRMSE} & SSIM & \multicolumn{1}{c|}{NRMSE} & SSIM \\ \hline
\multicolumn{1}{|c|}{\textbf{InD1}} & \multicolumn{1}{c|}{0.257} & 0.698 & \multicolumn{1}{c|}{0.240} & 0.711  & \multicolumn{1}{c|}{{\ul 0.126}} & {\ul 0.792} & \multicolumn{1}{c|}{\textbf{0.096}} & \textbf{0.869} \\ \hline
\multicolumn{1}{|c|}{\textbf{InD2}} & \multicolumn{1}{c|}{0.328} & 0.612 & \multicolumn{1}{c|}{0.319} & 0.641 & \multicolumn{1}{c|}{\textbf{0.238}} & \textbf{0.732} & \multicolumn{1}{c|}{{\ul 0.258}} & {\ul 0.724} \\ \hline
\multicolumn{1}{|c|}{\textbf{OOD}} & \multicolumn{1}{c|}{0.469} & 0.585 & \multicolumn{1}{c|}{{\ul 0.450}} & {\ul 0.632} & \multicolumn{1}{c|}{\textbf{0.435}} & \textbf{0.642} & \multicolumn{1}{c|}{0.455} & 0.630 \\ \hline
\end{tabular}
\end{table}

\subsubsection{Effect of atom number and output patch size}
\begin{figure}[ht]
    \centering
    \includegraphics[width=0.9\textwidth]{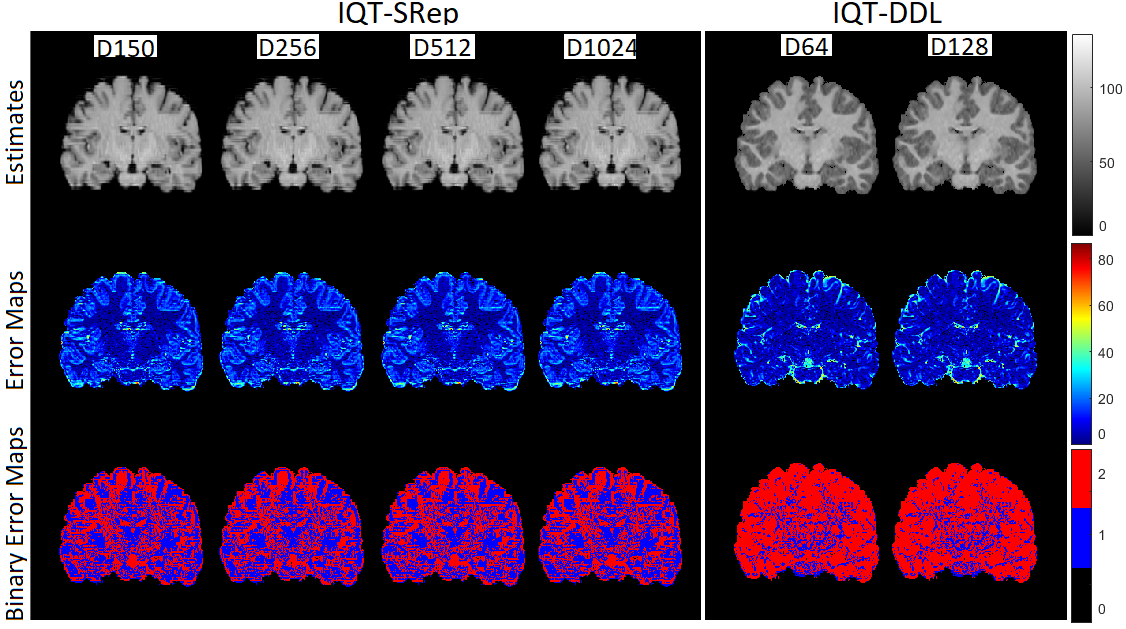}
    \caption{Effect of atom number for the IQT-SRep and IQT-DDL approaches using the OOD example in Fig. \ref{fig:Estimates}. Row 1: Image estimates using different atom numbers (from left to right: 150, 256, 512 and 1024 atoms (for IQT-SRep), and 64 and 128 atoms (for IQT-DDL). Row 2: Absolute difference error maps between the high quality image and each estimate in Row 1, and Row 3: Binary maps between interpolated low-field images and estimates in Row 1, where we can observe, as atom number increases, slightly more image regions (in red) are observed, until saturation from atom numbers of 512 to 1024 for IQT-SRep.}
    \label{fig:DictSize}
\end{figure}

Now, we evaluate the effect of atom number and patch size on both approaches. From the sampled 100,000 image patch pairs, and for the IQT-SRep approach, we train four dictionaries of size 150, 256, 512, 1024, and use each to estimate the high-field image from the low-field counterpart. Moreover, for the IQT-DDL approach, in order to assess the performance of the algorithm using different atoms numbers, we construct dictionaries using atom number of 64, in addition to that of 128 whose results are presented in the previous section. Table \ref{tab:NRMSEAtoms} shows NRMSE of the IQT-SRep and IQT-DDL approaches using different atom number using the three testing datasets InD1, InD2 and OOD. We can observe that in general, as atom number increases, construction quality improves (NRMSE decreases and SSIM increases), but saturates for atom number higher than 512 for the IQT-SRep approach. For the IQT-SRep approach, all tested atom numbers still provide better construction results using OOD data as compared to the IQT-DL approach. {On the other hand, we tested several patch sizes for both the IQT-SRep and IQT-DDL models. Specifically, for the IQT-SRep model, we tested patch sizes of P3: \( 3 \times 3 \times 3\), P5: \(5 \times 5 \times 5\), and P7: \(7 \times 7 \times 7\) using a dictionary size of 1024 (which provides the best results). For the IQT-DDL model, we tested patch sizes of P16: \(16 \times 16 \times 16\), P32: \(32 \times 32 \times 32\), and P48: \(48 \times 48 \times 48\) using a dictionary size of 128 (which provides the best results). Table \ref{tab:NRMSEPatch} provides the NRMSE and SSIM for two in-distribution datasets (InD1 and InD2) and one out-of-distribution (OOD) dataset. We can observe that for the IQT-SRep model, the reconstruction results improves (NRMSE decreases and SSIM increases) as the patch size increases. Conversely, for the IQT-DDL model, a patch size of P32: \(32 \times 32 \times 32\) outperforms both P16: \(16 \times 16 \times 16\) and P48: \(48 \times 48 \times 48\), indicating that it is a good operating point, balancing structural information content with the ability to learn and generalise from a finite training set.}
Fig. \ref{fig:DictSize} shows an example of super resolved images using the OOD data set using dictionaries of different sizes at patch sizes providing best results (P7 for IQT-SRep and P32 for IQT-DDL). While there are no substantial visual differences, we can observe in the binary error maps that the number red pixels (improvement over interpolated low-field image) gradually increase with larger dictionaries until saturation for dictionary size of 1024. In terms of computation time, the IQT-SRep algorithm is implemented in MATLAB and the experiments are carried out on a laptop with a 2.8 GHz processor CPU, with 16 GB of RAM, under Microsoft Windows 10. Dictionary construction times ranges from $\sim 25$ min for a 150-size dictionary to $\sim$ 80 min for a 1024-size dictionary. During the testing, in terms of test image reconstruction time, the computation is approximately linear to the size of the dictionary, that larger dictionaries will result in heavier computation. For example, smaller dictionaries, such as those with 150 atoms, yield reconstructions in an average time of $\sim 7$ min, while larger dictionaries, such as those with 1024 atoms, yielded image reconstructions in an average time of $50$ min. On the other hand, for the IQT-DDL algorithm, as shown in Table \ref{tab:NRMSEAtoms}, the NRMSE is slightly lower using dictionary with atom number of 128 compared to that of 64 atoms for all testing datasets InD1, InD2 and OOD, as it retains more image patterns. Fig. \ref{fig:DictSize} shows visual results of the IQT-DDL approach using an OOD example. Similar to the IQT-SRep approach, while there is no substantial visual difference, we indeed observe the increase in more super-resolved pixels (in red) in the binary error map images compared to the interpolated low-field image for atom number of 128 compared to that of 64. In terms of computation time, the IQT-DDL algorithm is implemented in PyTorch, and the testing construction time ranges from $\sim 4$ to 7 min for atom numbers of 64 to 128, respectively.
\begin{table}[]
\centering
\caption{{Effect of atom number for the IQT-SRep and IQT-DDL methods: NRMSE and SSIM using in-distribution data (InD1), and (InD2), and out-of-distribution data (OOD).}}
\label{tab:NRMSEAtoms}
\begin{tabular}{cc|cccc|cc|}
\cline{3-8}
\textbf{} &  & \multicolumn{4}{c|}{\textbf{IQT-SRep}} & \multicolumn{2}{c|}{\textbf{IQT-DDL}} \\ \cline{3-8} 
\textbf{} &  & \multicolumn{1}{c|}{\textbf{D150}} & \multicolumn{1}{c|}{\textbf{D256}} & \multicolumn{1}{c|}{\textbf{D512}} & \textbf{D1024} & \multicolumn{1}{c|}{\textbf{D64}} & \textbf{D128} \\ \hline
\multicolumn{1}{|c|}{\multirow{2}{*}{\textbf{InD1}}} & NRMSE & \multicolumn{1}{c|}{0.243} & \multicolumn{1}{c|}{0.242} & \multicolumn{1}{c|}{0.240} & 0.240 & \multicolumn{1}{c|}{0.128} & 0.126 \\ \cline{2-8} 
\multicolumn{1}{|c|}{} & SSIM & \multicolumn{1}{c|}{0.704} & \multicolumn{1}{c|}{0.705} & \multicolumn{1}{c|}{0.706} & 0.706   &  \multicolumn{1}{c|}{0.791} & 0.792 \\ \hline
\multicolumn{1}{|c|}{\multirow{2}{*}{\textbf{InD2}}} & NRMSE & \multicolumn{1}{c|}{0.322} & \multicolumn{1}{c|}{0.321} & \multicolumn{1}{c|}{0.319} & 0.319 & \multicolumn{1}{c|}{0.240} & 0.238 \\ \cline{2-8} 
\multicolumn{1}{|c|}{} & SSIM & \multicolumn{1}{c|}{0.639} & \multicolumn{1}{c|}{0.640} & \multicolumn{1}{c|}{0.641} & 0.641 & \multicolumn{1}{c|}{0.731} & 0.732 \\ \hline
\multicolumn{1}{|c|}{\multirow{2}{*}{\textbf{OOD}}} & NRMSE & \multicolumn{1}{c|}{0.452} & \multicolumn{1}{c|}{0.451} & \multicolumn{1}{c|}{0.450} & 0.450 & \multicolumn{1}{c|}{0.437} & 0.435 \\ \cline{2-8} 
\multicolumn{1}{|c|}{} & SSIM & \multicolumn{1}{c|}{0.630} & \multicolumn{1}{c|}{0.631} & \multicolumn{1}{c|}{0.632} & 0.632  & \multicolumn{1}{c|}{0.641} & 0.642 \\ \hline
\end{tabular}
\end{table}
\begin{table}[]
\centering
\caption{{Effect of output patch size for the IQT-SRep (using D1024) and IQT-DDL (using D128) methods: NRMSE and SSIM using in-distribution data (InD1), and (InD2), and out-of-distribution data (OOD).}}
\label{tab:NRMSEPatch}
\begin{tabular}{cc|ccc|ccc|}
\cline{3-8}
\textbf{} &  & \multicolumn{3}{c|}{\textbf{IQT-SRep}} & \multicolumn{3}{c|}{\textbf{IQT-DDL}} \\ \cline{3-8} 
\textbf{} &  & \multicolumn{1}{c|}{\textbf{P3}} & \multicolumn{1}{c|}{\textbf{P5}} & \textbf{P7} & \multicolumn{1}{c|}{\textbf{P16}} & \multicolumn{1}{c|}{\textbf{P32}} & \textbf{P48} \\ \hline
\multicolumn{1}{|c|}{\multirow{2}{*}{\textbf{InD1}}} & NRMSE & \multicolumn{1}{c|}{0.250} & \multicolumn{1}{c|}{0.244} & 0.240 & \multicolumn{1}{c|}{0.133} & \multicolumn{1}{c|}{0.126} & 0.129 \\ \cline{2-8} 
\multicolumn{1}{|c|}{} & SSIM & \multicolumn{1}{c|}{0.671} & \multicolumn{1}{c|}{0.702} & 0.706  & \multicolumn{1}{c|}{0.768} & \multicolumn{1}{c|}{0.792} & 0.785 \\ \hline
\multicolumn{1}{|c|}{\multirow{2}{*}{\textbf{InD2}}} & NRMSE & \multicolumn{1}{c|}{0.325} & \multicolumn{1}{c|}{0.322} & 0.319 & \multicolumn{1}{c|}{0.244} & \multicolumn{1}{c|}{0.238} & 0.237 \\ \cline{2-8} 
\multicolumn{1}{|c|}{} & SSIM & \multicolumn{1}{c|}{0.635} & \multicolumn{1}{c|}{0.639} & 0.641 & \multicolumn{1}{c|}{0.725} & \multicolumn{1}{c|}{0.732} &  0.729\\ \hline
\multicolumn{1}{|c|}{\multirow{2}{*}{\textbf{OOD}}} & NRMSE & \multicolumn{1}{c|}{0.459} & \multicolumn{1}{c|}{0.455} & 0.450 & \multicolumn{1}{c|}{0.440} & \multicolumn{1}{c|}{0.435} & 0.437 \\ \cline{2-8} 
\multicolumn{1}{|c|}{} & SSIM & \multicolumn{1}{c|}{0.625} & \multicolumn{1}{c|}{0.628} & 0.632 & \multicolumn{1}{c|}{0.635} & \multicolumn{1}{c|}{0.642} & 0.638 \\ \hline
\end{tabular}
\end{table}
\section{Discussion}
\label{sec:Discussion}
This work introduced two novel IQT approaches. To the best of our knowledge, it is the first time in the literature that an unsupervised learning and a blended supervised and unsupervised learning frameworks are considered for IQT. These approaches are introduced to highlight biased estimates that can result from supervised learning approaches, such as supervised deep learning, especially using out-of-distribution data. The main advantages of these two novel formulations are that they tend to avoid biased estimates using out-of-distribution test data, and are very well suited where there is a scarcity of training data. The first approach is based on a sparse representation and dictionary learning model, which trains two dictionaries using a sparse representation model from pairs of low- and high-quality volumes, whereas the second is based on a deep dictionary learning approach which explicitly learns high-resolution dictionary to upscale the input volume as in the sparse-coding-based methods, while the entire network, including high dictionary generator, is simultaneously optimised to take full advantage of deep learning methods. The performance of both approaches is demonstrated using a low-field MRI application, and compared against state-of-the-art supervised deep learning algorithm using both in-distribution and out-of-distribution datasets. Although supervised deep learning approach showed a superior performance using an in-distribution data set, one disadvantage of such class of methods is that their performance is degraded for images with a different contrast than in the training data set (OOD data). The results presented in the previous section show that the sparsity prior for image patches in the IQT-SRep and approach is effective in regularising the ill-posed IQT problem leading to good performance using out-of-distribution data compared to the IQT-DL approach. In these results, the dictionary size is fixed be 1024. Obviously, larger dictionaries retain more expressive patterns to the volumes of the trained data set, thus, yield more accurate approximation to the sparsity optimisation problem during the testing phase. However, this comes at the expense of increasing the computation cost. On the other hand, the blended learning IQT-DDL approach shows that the upsampling process is efficient because the main network does not need to maintain the information of the processed image at the pixel level in high-quality image space. Therefore, the network can concentrate only on predicting the coefficients of the high-quality dictionary yielding better performance using the InD2 and OOD datasets compared to IQT-DL and IQT-SRep. Extensive experiments show that sparse representation using dictionary learning and the deep dictionary learning approaches are more robust in super-resolving out-of-distribution test images compared to supervised deep learning. On the other hand, unsupervised learning is more robust to noise and redundancy in the data compared to supervised learning. Precisely, in a convex optimisation framework, training and testing samples are forced to follow the observation model of the imaging system on hand, and therefore, any new unseen test samples will follow this model, which can avoid the ``regression to the mean” problems observed with supervised regression models. The biased estimates produced using the IQT-DL model likely arise because these image regions are under-represented in the training data, and thus the model is under-fit, which further adds bias in estimates. Although other unsupervised approaches, in particular deep unsupervised learning, might produce better results, the proposed approach highlight the problem and provide a baseline potential solution. {It is worth pointing out that the proposed methods performed slightly better than the supervised approach only for data that were quite significantly different from the training dataset; this might be a not very relevant scenario for simulation-based training approaches, as most of the current IQT implementations are, as it would be much more advantageous to adapt the simulation parameters for the specific application and train an ad hoc supervised model than to adopt an unsupervised approach. However, there are several situations in which test images may have features that are difficult to simulate or predict, e.g. pathological alterations or artifacts. Some applications may also require very complex models that would be impractical to retrain for every single applications. In both these cases, it is critical to have a model that is robust enough to OOD data. Furthermore, as already mentioned, this was a proof-of-concept study considering relatively simple supervised approaches. More advanced methods will be investigated in the future and are expected to provide a more significant advantage compared to supervised baselines.}

\section{Conclusion and Future Work}
\label{sec:Conclusion}
In this work, we introduced two novel formulations of the IQT problem, which use an unsupervised learning framework, and a blended supervised and unsupervised learning, respectively. The unsupervised learning approach considers a sparse representation and dictionary learning model, whereas the combination of supervised and unsupervised learning approach is based on deep dictionary learning. The two models are evaluated using a low-field magnetic resonance imaging application aiming to recover high-quality images akin to those obtained from high-field scanners. Experiments comparing the proposed approaches against state-of-the-art deep learning IQT method identified that the two novel formulations of the IQT problem can avoid bias associated with supervised methods when tested using out-of-distribution data that differs from the distribution of the data the model was trained on. This highlights the potential benefit of these novel paradigms for IQT. Future work involves demonstrating the performance of the approach using real low-field MRI data and providing uncertainty bounds to the estimates, as well as extension to deep unsupervised methods that can combine the high fidelity appearance of supervised deep learning approaches to image enhancement with the reduced bias provided by unsupervised learning. The reduction of bias is an important step in the deployment of learning based methods for image enhancement, itself a vital component in the realisation of the potential of emerging low-field and portable MRI systems particularly for deployment in regions where accessibility is currently low. 


\acks{\noindent The 3T T1-weighted and T2-weighted images were provided in part by the Human Connectome Project, WU-Minn Consortium (Principal Investigators: David Van Essen and Kamil Ugurbil; 1U54MH091657) funded by the 16 NIH Institutes and Centers that support the NIH Blueprint for Neuroscience Research, United States; and by the McDonnell Center for Systems Neuroscience at Washington University, United States. For the purpose of open access, the author has applied a Creative Commons Attribution (CC BY) licence to any Author Accepted Manuscript version arising.}

%
\ethics{The work follows appropriate ethical standards in conducting research and writing the manuscript, following all applicable laws and regulations regarding treatment of animals or human subjects.}

\coi{\noindent The authors declare that they have no known competing financial interests or personal relationships that could have appeared to influence the work reported in this paper.}

\data{The IQT models are trained and evaluated on the publicly available high-field MRI datasets, i.e. the HCP data set for T1w and T2w images \cite{sotiropoulos2013advances}.}

\bibliography{main-melba}

\begin{thebibliography}{63}
\providecommand{\natexlab}[1]{#1}
\providecommand{\url}[1]{\texttt{#1}}
\expandafter\ifx\csname urlstyle\endcsname\relax
  \providecommand{\doi}[1]{doi: #1}\else
  \providecommand{\doi}{doi: \begingroup \urlstyle{rm}\Url}\fi

\bibitem[Alexander et~al.(2014)Alexander, Zikic, Zhang, Zhang, and Criminisi]{alexander2014image}
Daniel~C Alexander, Darko Zikic, Jiaying Zhang, Hui Zhang, and Antonio Criminisi.
\newblock Image quality transfer via random forest regression: applications in diffusion mri.
\newblock In \emph{International Conference on Medical Image Computing and Computer-Assisted Intervention}, pages 225--232. Springer, 2014.

\bibitem[Alexander et~al.(2017)Alexander, Zikic, Ghosh, Tanno, Wottschel, Zhang, Kaden, Dyrby, Sotiropoulos, Zhang, et~al.]{alexander2017image}
Daniel~C Alexander, Darko Zikic, Aurobrata Ghosh, Ryutaro Tanno, Viktor Wottschel, Jiaying Zhang, Enrico Kaden, Tim~B Dyrby, Stamatios~N Sotiropoulos, Hui Zhang, et~al.
\newblock Image quality transfer and applications in diffusion mri.
\newblock \emph{NeuroImage}, 152:\penalty0 283--298, 2017.

\bibitem[Anazodo et~al.(2022)Anazodo, Ng, Ehiogu, Obungoloch, Fatade, Mutsaerts, Secca, Diop, Opadele, Alexander, et~al.]{anazodo2022framework}
Udunna~C Anazodo, Jinggang~J Ng, Boaz Ehiogu, Johnes Obungoloch, Abiodun Fatade, Henk~JMM Mutsaerts, Mario~Forjaz Secca, Mamadou Diop, Abayomi Opadele, Dianel~C Alexander, et~al.
\newblock A framework for advancing sustainable mri access in africa.
\newblock \emph{medRxiv}, 2022.

\bibitem[Blumberg et~al.(2018)Blumberg, Tanno, Kokkinos, and Alexander]{blumberg2018deeper}
Stefano~B Blumberg, Ryutaro Tanno, Iasonas Kokkinos, and Daniel~C Alexander.
\newblock Deeper image quality transfer: Training low-memory neural networks for 3d images.
\newblock In \emph{International Conference on Medical Image Computing and Computer-Assisted Intervention}, pages 118--125. Springer, 2018.

\bibitem[Chen and Donoho(1994)]{chen1994basis}
Shaobing Chen and David Donoho.
\newblock Basis pursuit.
\newblock In \emph{Proceedings of 1994 28th Asilomar Conference on Signals, Systems and Computers}, volume~1, pages 41--44. IEEE, 1994.

\bibitem[Deshpande et~al.(2020)Deshpande, Chandra, and Balamurali]{deshpande2020deep}
S~Deshpande, M~Girish Chandra, and P~Balamurali.
\newblock Deep dictionary learning for inpainting.
\newblock In \emph{Computer Vision, Pattern Recognition, Image Processing, and Graphics: 7th National Conference, NCVPRIPG 2019, Hubballi, India, December 22--24, 2019, Revised Selected Papers}, volume 1249, page~79. Springer Nature, 2020.

\bibitem[Dong et~al.(2014)Dong, Loy, He, and Tang]{dong2014learning}
Chao Dong, Chen~Change Loy, Kaiming He, and Xiaoou Tang.
\newblock Learning a deep convolutional network for image super-resolution.
\newblock In \emph{Computer Vision--ECCV 2014: 13th European Conference, Zurich, Switzerland, September 6-12, 2014, Proceedings, Part IV 13}, pages 184--199. Springer, 2014.

\bibitem[Elad(2010)]{elad2010sparse}
Michael Elad.
\newblock \emph{Sparse and redundant representations: from theory to applications in signal and image processing}, volume~2.
\newblock Springer, 2010.

\bibitem[Elad and Aharon(2006)]{elad2006image}
Michael Elad and Michal Aharon.
\newblock Image denoising via learned dictionaries and sparse representation.
\newblock In \emph{2006 IEEE Computer Society Conference on Computer Vision and Pattern Recognition (CVPR'06)}, volume~1, pages 895--900. IEEE, 2006.

\bibitem[Eldaly(2018)]{eldalyBayesian}
Ahmed~Karam Eldaly.
\newblock \emph{Bayesian Image Restoration and Bacteria Detection in Optical Endomicroscopy}.
\newblock PhD thesis, Heriot-Watt University, Edinburgh, United Kingdom, 2018.

\bibitem[Eldaly and Alexander(2024)]{Eldaly2024Bayesian}
Ahmed~Karam Eldaly and Daniel~C. Alexander.
\newblock Bayesian magnetic resonance image reconstruction and uncertainty quantification.
\newblock In \emph{ISMRM \& SMRT Virtual Conference \& Exhibition}, Singapore, May 2024.

\bibitem[Eldaly et~al.(2019)Eldaly, Altmann, Akram, Perperidis, Dhaliwal, and McLaughlin]{eldaly2019patch}
Ahmed~Karam Eldaly, Yoann Altmann, Ahsan Akram, Antonios Perperidis, Kevin Dhaliwal, and Steve McLaughlin.
\newblock Patch-based sparse representation for bacterial detection.
\newblock In \emph{2019 IEEE 16th International Symposium on Biomedical Imaging (ISBI 2019)}, pages 657--661. IEEE, 2019.

\bibitem[Eldaly et~al.(2025)Eldaly, Figini, and Alexander]{Eldaly2024BayesianX}
Ahmed~Karam Eldaly, Matteo Figini, and Daniel Alexander.
\newblock Bayesian magnetic resonance joint image reconstruction and uncertainty quantification using sparsity prior models.
\newblock \emph{Submitted to IEEE transactions on Computational imaging}, 2025.

\bibitem[Gu et~al.(2019)Gu, Lu, Zuo, and Dong]{gu2019blind}
Jinjin Gu, Hannan Lu, Wangmeng Zuo, and Chao Dong.
\newblock Blind super-resolution with iterative kernel correction.
\newblock In \emph{Proceedings of the IEEE/CVF Conference on Computer Vision and Pattern Recognition}, pages 1604--1613, 2019.

\bibitem[Gyori et~al.(2022)Gyori, Palombo, Clark, Zhang, and Alexander]{gyori2022training}
Noemi~G Gyori, Marco Palombo, Christopher~A Clark, Hui Zhang, and Daniel~C Alexander.
\newblock Training data distribution significantly impacts the estimation of tissue microstructure with machine learning.
\newblock \emph{Magnetic resonance in medicine}, 87\penalty0 (2):\penalty0 932--947, 2022.

\bibitem[He et~al.(2015)He, Zhang, Ren, and Sun]{he2015delving}
Kaiming He, Xiangyu Zhang, Shaoqing Ren, and Jian Sun.
\newblock Delving deep into rectifiers: Surpassing human-level performance on imagenet classification.
\newblock In \emph{Proceedings of the IEEE international conference on computer vision}, pages 1026--1034, 2015.

\bibitem[Huang and Dragotti(2018)]{huang2018deep}
Jun-Jie Huang and Pier~Luigi Dragotti.
\newblock A deep dictionary model for image super-resolution.
\newblock In \emph{2018 IEEE International Conference on Acoustics, Speech and Signal Processing (ICASSP)}, pages 6777--6781. IEEE, 2018.

\bibitem[Iglesias et~al.(2023)Iglesias, Billot, Balbastre, Magdamo, Arnold, Das, Edlow, Alexander, Golland, and Fischl]{iglesias2023synthsr}
Juan~E Iglesias, Benjamin Billot, Ya{\"e}l Balbastre, Colin Magdamo, Steven~E Arnold, Sudeshna Das, Brian~L Edlow, Daniel~C Alexander, Polina Golland, and Bruce Fischl.
\newblock Synthsr: A public ai tool to turn heterogeneous clinical brain scans into high-resolution t1-weighted images for 3d morphometry.
\newblock \emph{Science advances}, 9\penalty0 (5):\penalty0 eadd3607, 2023.

\bibitem[Iglesias et~al.(2021)Iglesias, Billot, Balbastre, Tabari, Conklin, Gonz{\'a}lez, Alexander, Golland, Edlow, Fischl, et~al.]{iglesias2021joint}
Juan~Eugenio Iglesias, Benjamin Billot, Ya{\"e}l Balbastre, Azadeh Tabari, John Conklin, R~Gilberto Gonz{\'a}lez, Daniel~C Alexander, Polina Golland, Brian~L Edlow, Bruce Fischl, et~al.
\newblock Joint super-resolution and synthesis of 1 mm isotropic mp-rage volumes from clinical mri exams with scans of different orientation, resolution and contrast.
\newblock \emph{NeuroImage}, 237:\penalty0 118206, 2021.

\bibitem[Iglesias et~al.(2022)Iglesias, Schleicher, Laguna, Billot, Schaefer, McKaig, Goldstein, Sheth, Rosen, and Kimberly]{iglesias2022quantitative}
Juan~Eugenio Iglesias, Riana Schleicher, Sonia Laguna, Benjamin Billot, Pamela Schaefer, Brenna McKaig, Joshua~N Goldstein, Kevin~N Sheth, Matthew~S Rosen, and W~Taylor Kimberly.
\newblock Quantitative brain morphometry of portable low-field-strength mri using super-resolution machine learning.
\newblock \emph{Radiology}, 306\penalty0 (3):\penalty0 e220522, 2022.

\bibitem[Kim et~al.(2016)Kim, Lee, and Lee]{kim2016deeply}
Jiwon Kim, Jung~Kwon Lee, and Kyoung~Mu Lee.
\newblock Deeply-recursive convolutional network for image super-resolution.
\newblock In \emph{Proceedings of the IEEE conference on computer vision and pattern recognition}, pages 1637--1645, 2016.

\bibitem[Kim et~al.(2023)Kim, Tregidgo, Eldaly, Figini, and Alexander]{kim20233d}
Seunghoi Kim, Henry~FJ Tregidgo, Ahmed~K Eldaly, Matteo Figini, and Daniel~C Alexander.
\newblock A 3d conditional diffusion model for image quality transfer--an application to low-field mri.
\newblock \emph{arXiv preprint arXiv:2311.06631}, 2023.

\bibitem[Kingma and Ba(2014)]{kingma2014adam}
Diederik~P Kingma and Jimmy Ba.
\newblock {Adam: A method for stochastic optimization}.
\newblock \emph{arXiv preprint arXiv:1412.6980}, 2014.

\bibitem[Krizhevsky et~al.(2017)Krizhevsky, Sutskever, and Hinton]{krizhevsky2017imagenet}
Alex Krizhevsky, Ilya Sutskever, and Geoffrey~E Hinton.
\newblock Imagenet classification with deep convolutional neural networks.
\newblock \emph{Communications of the ACM}, 60\penalty0 (6):\penalty0 84--90, 2017.

\bibitem[Lau et~al.(2023)Lau, Xiao, Zhao, Su, Ding, Man, Wang, Tsang, Cao, Lau, et~al.]{lau2023pushing}
Vick Lau, Linfang Xiao, Yujiao Zhao, Shi Su, Ye~Ding, Christopher Man, Xunda Wang, Anderson Tsang, Peng Cao, Gary~KK Lau, et~al.
\newblock Pushing the limits of low-cost ultralow-field mri by dual-acquisition deep learning 3d superresolution.
\newblock \emph{Magnetic Resonance in Medicine}, 2023.

\bibitem[Li et~al.(2012)Li, Yin, and Fang]{li2012group}
Shutao Li, Haitao Yin, and Leyuan Fang.
\newblock Group-sparse representation with dictionary learning for medical image denoising and fusion.
\newblock \emph{IEEE Transactions on biomedical engineering}, 59\penalty0 (12):\penalty0 3450--3459, 2012.

\bibitem[Li et~al.(2024)Li, Yang, Fu, Yue, and Zhou]{li2024deep}
Yixiao Li, Xiaoyuan Yang, Jun Fu, Guanghui Yue, and Wei Zhou.
\newblock Deep bi-directional attention network for image super-resolution quality assessment.
\newblock \emph{arXiv preprint arXiv:2403.10406}, 2024.

\bibitem[Lim et~al.(2017)Lim, Son, Kim, Nah, and Mu~Lee]{lim2017enhanced}
Bee Lim, Sanghyun Son, Heewon Kim, Seungjun Nah, and Kyoung Mu~Lee.
\newblock Enhanced deep residual networks for single image super-resolution.
\newblock In \emph{Proceedings of the IEEE conference on computer vision and pattern recognition workshops}, pages 136--144, 2017.

\bibitem[Lin et~al.(2019)Lin, Figini, Tanno, Blumberg, Kaden, Ogbole, Brown, D’Arco, Carmichael, Lagunju, et~al.]{lin2019deep}
Hongxiang Lin, Matteo Figini, Ryutaro Tanno, Stefano~B Blumberg, Enrico Kaden, Godwin Ogbole, Biobele~J Brown, Felice D’Arco, David~W Carmichael, Ikeoluwa Lagunju, et~al.
\newblock Deep learning for low-field to high-field mr: image quality transfer with probabilistic decimation simulator.
\newblock In \emph{International Workshop on Machine Learning for Medical Image Reconstruction}, pages 58--70. Springer, 2019.

\bibitem[Lin et~al.(2021)Lin, Zhou, Slator, and Alexander]{lin2021generalised}
Hongxiang Lin, Yukun Zhou, Paddy~J Slator, and Daniel~C Alexander.
\newblock Generalised super resolution for quantitative mri using self-supervised mixture of experts.
\newblock In \emph{International Conference on Medical Image Computing and Computer-Assisted Intervention}, pages 44--54. Springer, 2021.

\bibitem[Lin et~al.(2022)Lin, Figini, D’Arco, Ogbole, Tanno, Blumberg, Ronan, Brown, Carmichael, Lagunju, Helen~Cross, Fernandez-Reyes, and Alexander]{Hongxiang2022}
Hongxiang Lin, Matteo Figini, Felice D’Arco, Godwin Ogbole, Ryutaro Tanno, Stefano Blumberg, Lisa Ronan, Biobele~J. Brown, David~W. Carmichael, Ikeoluwa Lagunju, Judith Helen~Cross, Delmiro Fernandez-Reyes, and Daniel~C. Alexander.
\newblock Low-field magnetic resonance image enhancement via stochastic image quality transfer.
\newblock \emph{Medical Image Analysis}, 2022.

\bibitem[Liu et~al.(2017)Liu, Zhai, Gu, Liu, Zhao, and Gao]{liu2017reduced}
Yutao Liu, Guangtao Zhai, Ke~Gu, Xianming Liu, Debin Zhao, and Wen Gao.
\newblock Reduced-reference image quality assessment in free-energy principle and sparse representation.
\newblock \emph{IEEE Transactions on Multimedia}, 20\penalty0 (2):\penalty0 379--391, 2017.

\bibitem[Liu et~al.(2018)Liu, Gu, Wang, Zhao, and Gao]{liu2018blind}
Yutao Liu, Ke~Gu, Shiqi Wang, Debin Zhao, and Wen Gao.
\newblock Blind quality assessment of camera images based on low-level and high-level statistical features.
\newblock \emph{IEEE Transactions on Multimedia}, 21\penalty0 (1):\penalty0 135--146, 2018.

\bibitem[Liu et~al.(2019)Liu, Gu, Zhang, Li, Zhai, Zhao, and Gao]{liu2019unsupervised}
Yutao Liu, Ke~Gu, Yongbing Zhang, Xiu Li, Guangtao Zhai, Debin Zhao, and Wen Gao.
\newblock Unsupervised blind image quality evaluation via statistical measurements of structure, naturalness, and perception.
\newblock \emph{IEEE Transactions on Circuits and Systems for Video Technology}, 30\penalty0 (4):\penalty0 929--943, 2019.

\bibitem[Liu et~al.(2024)Liu, Zhang, Hu, Gu, Zhai, and Dong]{liu2024underwater}
Yutao Liu, Baochao Zhang, Runze Hu, Ke~Gu, Guangtao Zhai, and Junyu Dong.
\newblock Underwater image quality assessment: Benchmark database and objective method.
\newblock \emph{IEEE Transactions on Multimedia}, 2024.

\bibitem[Loffe and Normalization(2014)]{loffe2014accelerating}
S~Loffe and CSB Normalization.
\newblock Accelerating deep network training by reducing internal covariate shift.
\newblock \emph{arXiv}, 2014.

\bibitem[Majumdar and Singhal(2017)]{majumdar2017noisy}
Angshul Majumdar and Vanika Singhal.
\newblock Noisy deep dictionary learning: Application to alzheimer's disease classification.
\newblock In \emph{2017 International Joint Conference on Neural Networks (IJCNN)}, pages 2679--2683. IEEE, 2017.

\bibitem[Majumdar and Ward(2017)]{majumdar2017robust}
Angshul Majumdar and Rabab Ward.
\newblock Robust greedy deep dictionary learning for ecg arrhythmia classification.
\newblock In \emph{2017 International Joint Conference on Neural Networks (IJCNN)}, pages 4400--4407. IEEE, 2017.

\bibitem[Manjani et~al.(2017)Manjani, Tariyal, Vatsa, Singh, and Majumdar]{manjani2017detecting}
Ishan Manjani, Snigdha Tariyal, Mayank Vatsa, Richa Singh, and Angshul Majumdar.
\newblock Detecting silicone mask-based presentation attack via deep dictionary learning.
\newblock \emph{IEEE Transactions on Information Forensics and Security}, 12\penalty0 (7):\penalty0 1713--1723, 2017.

\bibitem[Nair and Hinton(2010)]{nair2010rectified}
Vinod Nair and Geoffrey~E Hinton.
\newblock Rectified linear units improve restricted boltzmann machines.
\newblock In \emph{Proceedings of the 27th international conference on machine learning (ICML-10)}, pages 807--814, 2010.

\bibitem[Obermeyer et~al.(2019)Obermeyer, Powers, Vogeli, and Mullainathan]{obermeyer2019dissecting}
Ziad Obermeyer, Brian Powers, Christine Vogeli, and Sendhil Mullainathan.
\newblock Dissecting racial bias in an algorithm used to manage the health of populations.
\newblock \emph{Science}, 366\penalty0 (6464):\penalty0 447--453, 2019.

\bibitem[Sharma et~al.(2017)Sharma, Abrol, and Sao]{sharma2017deep}
Pulkit Sharma, Vinayak Abrol, and Anil~Kumar Sao.
\newblock Deep-sparse-representation-based features for speech recognition.
\newblock \emph{IEEE/ACM Transactions on Audio, Speech, and Language Processing}, 25\penalty0 (11):\penalty0 2162--2175, 2017.

\bibitem[Shocher et~al.(2018)Shocher, Cohen, and Irani]{shocher2018zero}
Assaf Shocher, Nadav Cohen, and Michal Irani.
\newblock “zero-shot” super-resolution using deep internal learning.
\newblock In \emph{Proceedings of the IEEE conference on computer vision and pattern recognition}, pages 3118--3126, 2018.

\bibitem[Singh and Majumdar(2017)]{singh2017deep}
Shikha Singh and Angshul Majumdar.
\newblock Deep sparse coding for non--intrusive load monitoring.
\newblock \emph{IEEE Transactions on Smart Grid}, 9\penalty0 (5):\penalty0 4669--4678, 2017.

\bibitem[Soh et~al.(2020)Soh, Cho, and Cho]{soh2020meta}
Jae~Woong Soh, Sunwoo Cho, and Nam~Ik Cho.
\newblock Meta-transfer learning for zero-shot super-resolution.
\newblock In \emph{Proceedings of the IEEE/CVF Conference on Computer Vision and Pattern Recognition}, pages 3516--3525, 2020.

\bibitem[Sotiropoulos et~al.(2013{\natexlab{a}})Sotiropoulos, Jbabdi, Xu, Andersson, Moeller, Auerbach, Glasser, Hernandez, Sapiro, Jenkinson, Feinberg, Yacoub, Lenglet, {Van Essen}, Ugurbil, and Behrens]{Sotiropoulos2013}
Stamatios~N. Sotiropoulos, Saad Jbabdi, Junqian Xu, Jesper~L. Andersson, Steen Moeller, Edward~J. Auerbach, Matthew~F. Glasser, Moises Hernandez, Guillermo Sapiro, Mark Jenkinson, David~A. Feinberg, Essa Yacoub, Christophe Lenglet, David~C. {Van Essen}, Kamil Ugurbil, and Timothy~E.J. Behrens.
\newblock {Advances in diffusion MRI acquisition and processing in the Human Connectome Project}.
\newblock \emph{NeuroImage}, 80:\penalty0 125--143, oct 2013{\natexlab{a}}.
\newblock ISSN 10538119.

\bibitem[Sotiropoulos et~al.(2013{\natexlab{b}})Sotiropoulos, Jbabdi, Xu, Andersson, Moeller, Auerbach, Glasser, Hernandez, Sapiro, Jenkinson, et~al.]{sotiropoulos2013advances}
Stamatios~N Sotiropoulos, Saad Jbabdi, Junqian Xu, Jesper~L Andersson, Steen Moeller, Edward~J Auerbach, Matthew~F Glasser, Moises Hernandez, Guillermo Sapiro, Mark Jenkinson, et~al.
\newblock Advances in diffusion mri acquisition and processing in the human connectome project.
\newblock \emph{Neuroimage}, 80:\penalty0 125--143, 2013{\natexlab{b}}.

\bibitem[Tai et~al.(2017)Tai, Yang, and Liu]{tai2017image}
Ying Tai, Jian Yang, and Xiaoming Liu.
\newblock Image super-resolution via deep recursive residual network.
\newblock In \emph{Proceedings of the IEEE conference on computer vision and pattern recognition}, pages 3147--3155, 2017.

\bibitem[Tang et~al.(2020)Tang, Liu, Xiao, and Sebe]{tang2020dictionary}
Hao Tang, Hong Liu, Wei Xiao, and Nicu Sebe.
\newblock When dictionary learning meets deep learning: Deep dictionary learning and coding network for image recognition with limited data.
\newblock \emph{IEEE transactions on neural networks and learning systems}, 32\penalty0 (5):\penalty0 2129--2141, 2020.

\bibitem[Tanno et~al.(2021)Tanno, Worrall, Kaden, Ghosh, Grussu, Bizzi, Sotiropoulos, Criminisi, and Alexander]{tanno2021uncertainty}
Ryutaro Tanno, Daniel~E Worrall, Enrico Kaden, Aurobrata Ghosh, Francesco Grussu, Alberto Bizzi, Stamatios~N Sotiropoulos, Antonio Criminisi, and Daniel~C Alexander.
\newblock Uncertainty modelling in deep learning for safer neuroimage enhancement: Demonstration in diffusion mri.
\newblock \emph{NeuroImage}, 225:\penalty0 117366, 2021.

\bibitem[Tariyal et~al.(2016)Tariyal, Majumdar, Singh, and Vatsa]{tariyal2016deep}
Snigdha Tariyal, Angshul Majumdar, Richa Singh, and Mayank Vatsa.
\newblock Deep dictionary learning.
\newblock \emph{IEEE Access}, 4:\penalty0 10096--10109, 2016.

\bibitem[Wang et~al.(2004)Wang, Bovik, Sheikh, and Simoncelli]{wang2004image}
Zhou Wang, Alan~C Bovik, Hamid~R Sheikh, and Eero~P Simoncelli.
\newblock Image quality assessment: from error visibility to structural similarity.
\newblock \emph{IEEE transactions on image processing}, 13\penalty0 (4):\penalty0 600--612, 2004.

\bibitem[Xu et~al.(2020)Xu, Tseng, Tseng, Kuo, and Tsai]{xu2020unified}
Yu-Syuan Xu, Shou-Yao~Roy Tseng, Yu~Tseng, Hsien-Kai Kuo, and Yi-Min Tsai.
\newblock Unified dynamic convolutional network for super-resolution with variational degradations.
\newblock In \emph{Proceedings of the IEEE/CVF Conference on Computer Vision and Pattern Recognition}, pages 12496--12505, 2020.

\bibitem[Yang et~al.(2010)Yang, Wright, Huang, and Ma]{yang2010image}
Jianchao Yang, John Wright, Thomas~S Huang, and Yi~Ma.
\newblock Image super-resolution via sparse representation.
\newblock \emph{IEEE transactions on image processing}, 19\penalty0 (11):\penalty0 2861--2873, 2010.

\bibitem[Zeyde et~al.(2010)Zeyde, Elad, and Protter]{zeyde2010single}
Roman Zeyde, Michael Elad, and Matan Protter.
\newblock On single image scale-up using sparse-representations.
\newblock In \emph{International conference on curves and surfaces}, pages 711--730. Springer, 2010.

\bibitem[Zhang et~al.(2014)Zhang, Zhao, and Gao]{zhang2014group}
Jian Zhang, Debin Zhao, and Wen Gao.
\newblock Group-based sparse representation for image restoration.
\newblock \emph{IEEE transactions on image processing}, 23\penalty0 (8):\penalty0 3336--3351, 2014.

\bibitem[Zhang et~al.(2018{\natexlab{a}})Zhang, Zuo, and Zhang]{zhang2018learning}
Kai Zhang, Wangmeng Zuo, and Lei Zhang.
\newblock Learning a single convolutional super-resolution network for multiple degradations.
\newblock In \emph{Proceedings of the IEEE conference on computer vision and pattern recognition}, pages 3262--3271, 2018{\natexlab{a}}.

\bibitem[Zhang et~al.(2018{\natexlab{b}})Zhang, Li, Li, Wang, Zhong, and Fu]{zhang2018image}
Yulun Zhang, Kunpeng Li, Kai Li, Lichen Wang, Bineng Zhong, and Yun Fu.
\newblock Image super-resolution using very deep residual channel attention networks.
\newblock In \emph{Proceedings of the European conference on computer vision (ECCV)}, pages 286--301, 2018{\natexlab{b}}.

\bibitem[Zhao et~al.(2017)Zhao, Sun, and Zhang]{zhao2017single}
Liling Zhao, Quansen Sun, and Zelin Zhang.
\newblock Single image super-resolution based on deep learning features and dictionary model.
\newblock \emph{IEEE Access}, 5:\penalty0 17126--17135, 2017.

\bibitem[Zhou and Susstrunk(2019)]{zhou2019kernel}
Ruofan Zhou and Sabine Susstrunk.
\newblock Kernel modeling super-resolution on real low-resolution images.
\newblock In \emph{Proceedings of the IEEE/CVF International Conference on Computer Vision}, pages 2433--2443, 2019.

\bibitem[Zhou et~al.(2020)Zhou, Jiang, Wang, Chen, and Li]{zhou2020blind}
Wei Zhou, Qiuping Jiang, Yuwang Wang, Zhibo Chen, and Weiping Li.
\newblock Blind quality assessment for image superresolution using deep two-stream convolutional networks.
\newblock \emph{Information Sciences}, 528:\penalty0 205--218, 2020.

\bibitem[Zhou et~al.(2021)Zhou, Wang, and Chen]{zhou2021image}
Wei Zhou, Zhou Wang, and Zhibo Chen.
\newblock Image super-resolution quality assessment: Structural fidelity versus statistical naturalness.
\newblock In \emph{2021 13th International conference on quality of multimedia experience (QoMEX)}, pages 61--64. IEEE, 2021.

\bibitem[Zhou et~al.(2018)Zhou, Rahman~Siddiquee, Tajbakhsh, and Liang]{zhou2018unet++}
Zongwei Zhou, Md~Mahfuzur Rahman~Siddiquee, Nima Tajbakhsh, and Jianming Liang.
\newblock Unet++: A nested u-net architecture for medical image segmentation.
\newblock In \emph{Deep Learning in Medical Image Analysis and Multimodal Learning for Clinical Decision Support: 4th International Workshop, DLMIA 2018, and 8th International Workshop, ML-CDS 2018, Held in Conjunction with MICCAI 2018, Granada, Spain, September 20, 2018, Proceedings 4}, pages 3--11. Springer, 2018.

\end{thebibliography}

\end{document}